\documentclass[11pt,a4paper]{article}
\usepackage[body={16.5cm,25cm}]{geometry}
\usepackage[dvips]{color}
\usepackage{array}
\usepackage{stackrel}
\usepackage{tensor}
\usepackage{mathtools}
\usepackage{curves}
\usepackage[titletoc]{appendix}

\usepackage[hang,small,center]{caption}
\usepackage{amsmath,amssymb}
\usepackage{amsfonts}
\usepackage{mathrsfs}
\usepackage{epic}
\usepackage{eepic}
\usepackage{graphicx}
\usepackage{cite}
\newcommand{\ds}{\displaystyle}

\renewcommand{\author}[1]{\large\rm #1\\ \bigskip}
\newcommand{\address}[1]{{\normalsize\it #1\\}\bigskip}
\renewcommand{\title}[1]{\bigskip\bigskip\Large\bf #1\bigskip\bigskip\\}

\newcommand{\Bigpsi}[3]{\phantom{\Psi}_2 \kern -.05em
\Psi_2\left(\genfrac{}{}{0pt}{}{#1}{#2}\biggl|#3\right)}

\def\bea{\begin{eqnarray}}
\def\eea{\end{eqnarray}}
\newcommand{\be}{\begin{equation}}
\newcommand{\ee}{\end{equation}}
\newcommand{\beq}{\begin{equation}}
\newcommand{\eeq}{\end{equation}}

\newcommand{\Ru}{\boldsymbol{ R}}

\newcommand{\bosn}{\textrm{\scriptsize $\boldsymbol{N}$}}

\newcommand{\lam}{{\lambda}}
\newcommand{\ip}{{i'_1}}
\newcommand{\jp}{{j'_1}}
\newcommand{\Rc}{{\cal{R}}}
\newcommand{\phif}{\tensor*[_{4}]{\overline{\phi}}{_3}}
\newcommand{\phifs}{\tensor*[_{4}]{{\phi}}{_3}}
\newcommand{\phit}{\tensor*[_{3}]{\overline{\phi}}{_2}}
\newcommand{\phis}{\tensor*[_{2}]{\overline{\phi}}{_1}}
\newcommand{\phiss}{\tensor*[_{2}]{{\phi}}{_1}}
\newcommand{\phir}{\tensor*[_{r+1}]{\overline{\phi}}{_r}}
\newcommand{\phirs}{\tensor*[_{r+1}]{{\phi}}{_r}}

\newcommand{\bos}{\boldsymbol{a}}

\newcounter{app}
\newcounter{sapp}[app]

\def\nsection#1{\setcounter{equation}{0}\section{#1}}

\begin{document}

\vglue 2 cm
\begin{center}
\title{On the Yang-Baxter equation for the
six-vertex model}
\author{Vladimir
  V.~Mangazeev$^{1,2}$ }
\address{$^1$Department of Theoretical Physics,
         Research School of Physics and Engineering,\\
    Australian National University, Canberra, ACT 0200, Australia.\\\ \\
$^2$Mathematical Sciences Institute,\\
      Australian National University, Canberra, ACT 0200,
      Australia.
}

\begin{abstract}
In this paper we review the theory of the Yang-Baxter equation related to
the 6-vertex model and its higher spin generalizations. We employ a 3D
approach to the problem.
Starting with the 3D R-matrix, we consider a two-layer projection of the
corresponding 3D lattice model. As a result, we obtain a new expression
for the higher spin $R$-matrix associated with the affine quantum algebra
$U_q(\widehat{sl(2)})$. In the simplest case of the spin
$s=1/2$ this $R$-matrix naturally reduces to the $R$-matrix of the 6-vertex model.
Taking a special limit in our construction we also obtain new formulas
for  the $Q$-operators acting in the representation space of arbitrary (half-)integer
spin. Remarkably, this construction can be naturally extended to any complex values
of spin $s$. We also give
all functional equations satisfied by the transfer-matrices and  $Q$-operators.
\end{abstract}

\end{center}

\newpage

\nsection{Introduction}

In this paper we  analyze the properties of the six-vertex model in an external field and its
higher spin generalizations based on a new 3D approach developed in
\cite{Bazhanov:2005as,Bazhanov:2008rd,MBS13}. This approach allows us to reveal new algebraic
and analytic properties of the six-vertex model with arbitrary
spin.

The theory of the six-vertex model goes back to the works of Lieb
\cite{Lieb66,Lieb67} who solved the famous two-dimensional ice model.
These results were further extended by Sutherland \cite{Sut67} to
the zero field six-vertex model and then generalized to the case of
an arbitrary electric field by Yang, Sutherland et al.
\cite{Yang1967zz, SYY67}. The main technique used was
the Bethe ansatz \cite{Bethe:1931hc}. However, in \cite{Bax72,Bax73a,Bax73b,Bax73c}
Baxter introduced new analytic and algebraic methods which allowed him to solve
the eight-vertex model in a zero field.
The main ingredient of Baxter's approach is the theory of functional equations based
on the concept
of the  $Q$-operator. This $Q$-operator satisfies the so-called $TQ$-relation
which allows in principle to calculate  eigenvalues of the transfer-matrix of the model.
Starting with the work \cite{Str79}, the analytic Bethe ansatz \cite{Res83}
was developed where the $TQ$-relation (or an analogous equation) is used as a formal
substitution to solve the transfer matrix functional equations.

The theory of functional relations allows us to determine  eigenvalues of the higher
transfer matrices associated with the so-called fusion procedure.
This algebraic procedure provides a derivation of the functional relations for the higher
transfer matrices based on decomposition properties of products of representations of
the affine quantum groups. The notion of ``higher'' spin (or ``fused'') $R$-matrices
was developed in \cite{KRS81}
from the point of view of representation theory. These $R$-matrices for the six-vertex model
acting in the tensor product of two highest weight modules were calculated in \cite{KR87a}.
However, the formulas derived in \cite{KR87a} involve special projection operators
and are not very convenient for practical calculations.

An alternative method for calculating  the higher spin $R$-matrices was developed by Jimbo
\cite{Jim85} (see also \cite{Del94} for all simple Lie algebras).
It is based on the spectral decomposition of the $R$-matrix
and allows one to calculate the $R$-matrix in terms of spectral functions and quantum
Clebsch-Gordan coefficients.
For example, in the $U_q(sl(2))$ case, it results in a triple sum formula for
the matrix elements
of the ``fused'' $R$-matrices.

The main result of this paper is a new representation of the
$U_q(sl(2))$ $R$-matrix $R_{I,J}(\lambda)$  in the
tensor product of two highest weight representations  with arbitrary weights $I$ and $J$.
It contains only one single summation and is expressed in terms of the basic hypergeometric series.
The explicit formula reads

\beq
[R_{I,J}(\lam;\phi)]_{i,j}^{i',j'}=\delta_{i+j,i'+j'}\,
\rho_{I,J}(\lambda)\,\phi^{2i}\,
a_{ij}^{i'j'}(\lambda)
\phif\left(\left.\begin{array}{l}
q^{-2i};q^{-2i'},\lambda^{-2} q^{J-I},\lambda^{2}q^{2+J-I}\\
\phantom{aa}q^{-2I},q^{2(1+j-i')},q^{2(1+J-i-j)}\end{array}\right|q^2,q^2\right)\label{intro1}
\eeq
with
\beq
a_{ij}^{i'j'}(\lambda)=(-1)^{i}\frac{q^{i(i+j-2J-1)-jI+i'(I+j')}}
{{\lambda^{i+i'}}(q^2;q^2)_i}
\frac{(q^{-2J};q^2)_j(\lambda^{-2}q^{I-J};q^2)_{j-i'}}
{(q^{-2J};q^2)_{j'}(\lambda^{-2}q^{-I-J};q^2)_{i+j}},\label{intro3}
\eeq
\noindent where $\rho_{I,J}(\lambda)$ is the
normalization factor and we defined  a {\it regularized} terminating
basic hypergeometric series $\phir$ as
\beq
\ds\phir(q^{-n};\{a\}_r;\{b\}_r|\,q,z)=
\sum_{k=0}^n z^k\,\frac{(q^{-n};q)_k}{(q;q)_k}\prod_{s=1}^r(a_s;q)_k (b_sq^k;q)_{n-k}\,.
\label{phi-reg0}
\eeq

It is easy to see that the hypergeometric series $\phif$ entering (\ref{intro1})
can be expressed in terms of the $q$-Racah polynomials \cite{AW79}.
The fact that $q$-Racah polynomials satisfy the Yang-Baxter equation
with a spectral parameter is quite remarkable and should have some profound origins.

Another important property of the $R$-matrix (\ref{intro1}) is that all its nonzero matrix elements
can be made positive under the proper choice of the spectral parameter $\lambda$ and
the normalization factor $\rho_{I,J}(\lambda)$. This is explained
in Section 4.

We notice that a similar formula with only one summation exists for the XXX spin chain in
a holomorphic basis
(see formula (2.17) in \cite{Skl85}). It would be interesting to understand its connection
with (\ref{intro1}).

As an application of the formula (\ref{intro1}) we construct the  $Q$-operators
related to the $U_q(\widehat{sl(2)})$ algebra as  special
transfer matrices  acting in the tensor product of arbitrary highest weight representations.
The idea of the construction of the $Q$-operator in terms of some special
transfer matrices belongs
to Baxter \cite{Bax72}. It is a key element of his original solution
 of the 8-vertex model. For the simplest case of the six-vertex model the quantum space is
built from 2-dimensional
highest weight representations of the $U_q(sl(2))$ algebra at every site of the lattice.

The next step in a better understanding of the structure of the $Q$-operators related
to the six-vertex model was achieved by Bazhanov and Stroganov \cite{BS90}. They considered
fundamental $L$-operators \cite{Faddeev:1979} intertwined by the $R$-matrix of
the six-vertex model at the roots of unity $q^N=1$. In this case,
 the highest weight representation of the  $U_q(sl(2))$ algebra is replaced
with a cyclic representation. Then all matrix elements of the
$Q$-operator can be explicitly calculated  as simple products involving only
a two-spin interaction. Remarkably, these $Q$-operators coincide with
the transfer matrix of the chiral Potts model \cite{vG85,AuY87,BPY87}.

A seemingly different method was developed by Pasquier and Gaudin \cite{PG92} where
they constructed
the $Q$-operator for the Toda lattice in the form of integral operator.
Their $Q$-operator has a factorized kernel and its quasi-classical asymptotics
gives a generating function for Backlund transformations in the corresponding classical system.
It appears that this construction  is naturally connected to a separation of
variables (SoV) in  quantum and classical integrable systems \cite{Sk95}.
Later on this approach has been successfully applied to many other quantum lattice
integrable systems
and the general scheme of quantum SoV has been developed \cite{KS98a, Skl00,KSS00,KMS03}.
The integral $Q$-operator for the case of the XXX chain was first calculated in \cite{Der99}.
It is worth noting that taking the limit $N\to\infty$ \cite{AuYang99} in the Bazhanov and Stroganov
construction \cite{BS90}
one can recover the results of \cite{PG92} and \cite{Der99}.

The main difference of the  above approach from the original Baxter method
is that the ``quantum'' representation space is infinite-dimensional.
It has the structure of a tensor product of Verma modules
with  the basis  chosen as multi-variable polynomials $p(x_1,\ldots,x_M)$,
where $M$ is the size of the system. The $Q$-operators appear as integral operators
with an explicit action on such a polynomial basis.
A detailed construction can be found in
 \cite{CDKK12a,CDKK12b} for the XXX case and its generalization to the XXZ case in
\cite{Der05,CDKK13}.
The non-compact case and applications of the $Q$-operators to Liouville theory are discussed in \cite{
Kas01,Bytsko:2006,Bytsko09}.
It is worth mentioning that the representation of the
$Q$-operator by an integral operator is known only for the XXX case \cite{Der99}.
The proper deformation of such integral operator for the case of the six-vertex model is still
a challenging problem.

Another problem arises when spins take (half-) integer values.
In this case the quantum space  becomes reducible and
the action of the $Q$-operator on the polynomial basis becomes singular. This difficulty
can be overcome by expanding near the limit $2s\to\mathbb{Z}_+$
as shown in \cite{CDKK12a,CDKK12b}.
However, a removal of such a regularization is technically challenging and
it is desirable to have an alternative approach which is free from this difficulty.

In 1997 Bazhanov, Lukyanov and Zamolodchikov (BLZ) suggested another method to derive the
$Q$-operators related to the affine algebra $U_q(\widehat{sl(2)})$
\cite{BLZ97a,BLZ99a}. Based on the universal $R$-matrix theory
\cite{Drinfeld1985} they showed that the $Q$-operators can be constructed as  special
monodromy operators with the auxiliary space being an infinite-dimensional representation
of the $q$-oscillator algebra.
Although their original approach was developed
in the context of quantum field theory, the results of  \cite{BLZ97a,BLZ99a}
can be easily adjusted to the spin $s=1/2$ XXZ chain \cite{BLZunpub}.
However, the derivation of the local $Q$-operators  from the universal
$R$-matrix \cite{TK92} quickly becomes unbearable for higher spins
and has been completed only for the $s=1/2$ case \cite{BLZunpub,BJMS06,Boos2010, BLMS10}.
In principle, one can use the fusion
procedure to derive the $Q$-operators in any highest weight representation
with $2s\in\mathbb{Z}_+$, but this is also technically challenging.

The original motivation of this work was to understand a connection between
the integral $Q$-operators with factorized kernels which appear in the case of
infinite-dimensional
representations (or the cyclic case $q^N=1$) and the BLZ construction.
As the first step, we need to calculate the local $Q$-operators acting in the tensor
product of the $q$-oscillator algebra and the highest weight representation with
the arbitrary weight $2s$.

It is  known \cite{BLZ97a,BLZ99a,Pronko99,YNZ05,Boos2013} that one can construct  XXZ (or XXX)
$Q$-operators by taking the infinite spin limit in the auxiliary space
of the higher spin transfer-matrix.
Our new formula
(\ref{intro1}) suits  this purpose perfectly.
We also notice here that a 3D approach we employ in this paper is useful for constructing universal
$R$-matrices for higher rank algebras \cite{KV00}.

Taking the limit  $I\to\infty$ in (\ref{intro1}) we derive a generalization of the  BLZ
$Q$-operators
acting in the tensor product of the highest weight Verma modules with the arbitrary weight
$J\in\mathbb{C}$.
The limit $J\to\mathbb{Z}_+$ is non-singular and gives
the $Q$-operator in any finite-dimensional representation. The case $J=1$  reduces
to the previously known BLZ $Q$-operators.

The paper is organized as follows. In Section 2 we
recall some basic facts about the $U_q(sl(2))$ and the $q$-oscillator algebras.
We also give a definition of the terminating basic hypergeometric series and
their regularized version, which we use in the paper.
In Section 3 we define the
3D $R$-matrix and discuss its basic properties following \cite{MBS13}.
In Section 4 we consider a two-layer projection and derive a formula for
the matrix elements of the
$U_q(sl(2))$ $R$-matrix $R_{I,J}(\lambda)$ acting in the tensor product
of two highest weight representations with integer weights $I$ and $J$.
In Section 4 we discuss
the properties of this $R$-matrix and show that for the case $I=1$ it reduces
to the standard $U_q(sl(2))$ $L$-operator acting in the $(J+1)$- dimensional
representation space.
Then we transform the formula for the $R$-matrix from Section 2 to a remarkably
simple formula (\ref{intro1}) which contains only one summation and can be rewritten
as a terminating
balanced $\phifs$ series.  Using this representation we prove two important symmetry
relations for
the $R$-matrix $R_{I,J}(\lambda)$.
We show that this construction can be generalized to
the case of infinite-dimensional highest weight representations with $I,J\in\mathbb{C}$.
In Section 5 we introduce two $Q$-operators ${\bf Q}_\pm(\lambda)$
acting in the tensor product of the highest weight modules
with the weight $I$. Based on the factorization property for the transfer-matrix
\cite{BLZ97a,BLZ99a} we derive explicit expressions for the
matrix elements of the local $Q$-operators for any values of $I$ (including the
infinite-dimensional case $I\in\mathbb{C}$).
We also derive the standard $TQ$-relation   and calculate the Wronskian
of its two solutions ${\bf Q}_\pm(\lambda)$.
In Section 6 we list a standard set of functional relations satisfied
by higher-spin transfer matrices and $Q$-operators.
Finally, in the Conclusion we summarize all results and outline further directions
of research.

\nsection{Conventions}

First, let us recall some simple facts about the $U_q(sl(2))$ algebra.
It is generated by three elements $E$, $F$ and $H$ with defining relations
\beq
q^{H}Eq^{-H}=q^2E,\quad q^{H}Fq^{-H}=q^{-2}F,\quad [E,F]=\frac{[q^H]}{[q]}\label{conv1}
\eeq
and the following Casimir element
\beq
C=[q]^2FE+\{q^{H+1}\}=[q]^2EF+\{q^{H-1}\},\label{conv3}
\eeq
where we used the following notations
\beq
[x]=x-x^{-1},\quad  \{x\}=x+x^{-1}.\label{conv2}
\eeq
The Casimir element is normally
parameterized by a complex number $J\in\mathbb{C}$
\beq
C=\{q^{J+1}\}.\label{conv4}
\eeq
For any $J\in\mathbb{C}$ one can introduce an infinite-dimensional Verma module
$V_J^+$ with a basis $v_j$, $j\in\mathbb{Z}_{+}$. We define the
infinite-dimensional representation $\pi_J^+$ of $U_q(sl(2))$
by the following action on the module $V_J^+$
\beq
H v_j=(J-2j)v_j,\quad E v_j=\frac{[q^j]}{[q]}v_{j-1},\quad
F v_j=\frac{[q^{J-j}]}{[q]}v_{j+1}.
\label{conv5}
\eeq
When $J\in\mathbb{Z_+}$, the representation $\pi_J^+$ becomes reducible.
The vectors $v_j$, $j>J$ span an irreducible submodule of $V^+_J$ isomorphic
to $V^+_{-J-2}$ and one can introduce a finite-dimensional module
$V_J$ with the basis $\{v_0,\ldots,v_J\}$ isomorphic to
the quotient module $V_J^+/V_{-J-2}^+$. We denote the corresponding
finite-dimensional representation
as $\pi_J$.

Now let us consider the $q$-oscillator algebra
\beq\label{q-osc1}
\mathsf{Osc}_{\,q}:\qquad
q^{\boldsymbol N}\bos^{\pm}=q^{\pm 1}\,\bos^{\pm}\,\;q^{\boldsymbol N}\qquad
q\,\bos^+\bos^- - q^{-1} \,\bos^-\bos^+=q-q^{-1},\qquad
\eeq
generated by three elements
${\boldsymbol N}$, $\bos^+$ and $\bos^-$
and impose an additional relation
\beq
q^{2{\boldsymbol N}}=(1 -\bos^+\bos^-)\equiv q^{-2}\,(1-\bos^-\bos^+)\,.
\label{q-osc2}
\eeq

To make a link with the $3D$ $R$-matrix from the next section we shall
introduce an infinite-dimensional Fock space $\mathcal{F}_q$,
 spanned by a set of vectors
$|n\rangle$, $n=0,1,2,\ldots,\infty$,  with the natural scalar product
\beq\label{N-def}
\langle m| n\rangle= \delta_{m,n}\,,\qquad {\boldsymbol N}\,
|n\rangle= n\, |n\rangle\,,\qquad
\langle n| \, {\boldsymbol N} =\langle n| \, n\,.
\eeq
 The algebra \eqref{q-osc1} has two irreducible
highest weight representations  on the space $\mathcal{F}_q$
which we denoted as ${\cal F}_q^\pm$ in \cite{MBS13}.

In this paper we shall only use one representation $\mathcal{F}_q^+$
with a slightly modified action comparing to \cite{MBS13}
\beq
\begin{array}{lll}
\bos^-|0\rangle=0,\quad &\bos^+|n\rangle=|n+1\rangle,\qquad
&\bos^-|n\rangle=(1-q^{2n})|n-1\rangle,
\\[.4cm]
\langle 0|\bos^+=0\,,\qquad& \langle n|\bos^+=\langle n-1|,
&\langle n | \bos^-=\langle n+1|(1-q^{2+2n})\,,\label{conv6}
\end{array}
\eeq
with $n=0,1,2\ldots$.

Now we need to define the trace operation over the representations of
$\pi_j^+$ and $\mathcal{F}_q$. Consider an operator ${\bf T}$ acting
in the tensor product $V_J^+\otimes W$, where $W$ is some ``quantum''
representation space of the $U_q(sl(2))$ algebra.
Then we define the trace ${\bf T}_J$ over $V_J^+$ of the operator ${\bf T}$ simply
summing over all $j$
\beq
{\bf T}_J=\underset{{ \,V^+_J}}{\mbox{ Tr}}\,({\bf T}).\label{conv7}
\eeq
The trace over the finite-dimensional representation $\pi_J$ is defined
in a similar way.

Now let us consider  an operator ${\bf A}(\phi)$ acting in the tensor product
$\mathcal{F}_q\otimes W$, where $\phi\in\mathbb{C}$ is a ``horizontal'' field.
We define a normalized trace
of  ${\bf A}(\phi)$
on the space ${\mathcal F}_q$
by
\beq
\underset{{\,\mathcal{F}_q}}{\widehat{\mbox{Tr}}}\,({\bf A(\phi)})=
\frac{\underset{{\,\mathcal{F}_q}}{{\mbox{Tr}}}\,({\bf A(\phi)})}
{\underset{{\,\mathcal{F}_q}}{{\mbox{Tr}}}\,
({\phi^{2{\boldsymbol N}} q^{-{{\boldsymbol N}}\otimes H}})},\label{conv8}
\eeq
where $H$ is the generator of the $U_q(sl(2))$ algebra acting in the quantum
space $W$. We always assume that the field variable $\phi$ is chosen
in such a way that corresponding geometric series converge and then analytically
continue to any values of $\phi$. We also notice a relation between the
field $\phi$ and the additive field $h$
\beq
\phi=q^{h}.\label{conv9}
\eeq
In this paper  we prefer to use the exponential field $\phi$.

In the last part of this section we remind a definition of the basic hypergeometric series
\cite{Gasper} which we use
in the next sections.
We start with a $q$-Pochhammer symbol
\beq
(x\,;q)_n=\prod_{k=0}^{n-1} (1-x\, q^{k}),\quad n\geq0\label{q-Poc}
\eeq
and
\beq
(x\,;q)_n=\frac{1}{(xq^n;q)_{-n}}=
\frac{q^{n(n+1)/2}(-x/q)^n}{(q/x;q)_{-n}},\quad n<0\,.\label{q-Poc1}
\eeq

In this paper we  consider only terminating basic hypergeometric series
 $\phirs$
which is defined by
\beq
\phirs(q^{-n},\{a\}_r;\{b\}_r|\,q,z)\equiv\phirs\left(\left.\begin{array}
{l}q^{-n},a_1,\ldots,a_r\\\phantom{q^{-n},}b_1,\ldots,b_r\end{array}\right|q,z\right)=
\sum_{k=0}^n z^k\,\frac{(q^{-n};q)_k}{(q;q)_k}
\prod_{s=1}^r\frac{(a_s;q)_k}{(b_s;q)_k}\, .\label{phi-gen}
\eeq
The formula (\ref{phi-gen}) is well defined for all $a_i,b_i\in\mathbb{C}$ except

the case when some
of the parameters $b_i$ are equal to non-positive integer powers of $q$, i.e. $b_i=q^{-n}$,
$n\in\mathbb{Z}_+$ for some $i$.
To overcome this restriction we shall also introduce
a {\it regularized} version of terminating basic hypergeometric series.
Unlike usual hypergeometric functions $\phantom{|}_{r+1}F_r$ there is no a commonly accepted
definition for { regularized} basic hypergeometric series.
So we find it convenient to define a  regularized terminating basic hypergeometric series $\phir$
 as
\begin{align}
\ds&\phir(q^{-n};\{a\}_r;\{b\}_r|\,q,z)\equiv
\phir\left(\left.\begin{array}
{l}q^{-n};a_1,\ldots,a_r\\
\phantom{q^{-n},}b_1,\ldots,b_r\end{array}\right|q,z\right)=\nonumber\\
 =&\phirs(q^{-n},\{a\}_r;\{b\}_r|\,q,z)\times\prod_{s=1}^r(b_s;q)_n
\ds =\sum_{k=0}^n z^k\,\frac{(q^{-n};q)_k}{(q;q)_k}\prod_{s=1}^r(a_s;q)_k (b_sq^k;q)_{n-k}\,.
\label{phi-reg}
\end{align}
The formula (\ref{phi-reg}) is obviously well defined for any $a_i,b_i\in\mathbb{C}$.

We also notice that the symmetry between $q^{-n}$ and $a_1,\ldots,a_r$ is  broken
and this is why we used an extra semicolon after the first argument of $\phir$ in (\ref{phi-reg}).

\nsection{The 3D $R$-matrix}

In \cite{MBS13} we defined the 3D $R$-matrix as the
operator $\boldsymbol{R}$ acting in the tensor product
of three Fock spaces $\mathcal{F}_q\otimes \mathcal{F}_q\otimes
\mathcal{F}_q$.
If we define states in $\mathcal{F}_q\otimes \mathcal{F}_q\otimes
\mathcal{F}_q$ as $|n_1,n_2,n_3\rangle=|n_1\rangle\otimes|n_2\rangle\otimes
|n_3\rangle$, then the operator $\boldsymbol{R}$ is completely determined
by its matrix elements
\beq\label{mat-def}
R_{\,n_1,\,n_2,\,n_3}^{{\,n'_1,\,n'_2,\,n'_3}^{\phantom{|}}}=
\langle n_1,n_2,n_3\,|\,\boldsymbol{R}\,|\,n'_1,n'_2,n'_3\rangle,\qquad
n_i,n'_i=0,1,2,\ldots\infty,\quad i=1,2,3,
\eeq
 where
 \beq
R_{\,n_1,\,n_2,\,n_3}^{{\,n'_1,\,n'_2,\,n'_3}^{\phantom{|}}}=
\delta_{n_1+n_2,n_1'+n_2'}\delta_{n_2+n_3,n_2'+n_3'}
\frac{q^{n_2(n_2+1)-(n_2-n_1')(n_2-n_3')}}
{(q^2;q^2)_{n_2}}
Q_{n_2}(q^{-2n_1'},q^{-2n_2'},q^{-2n_3'}),
\label{rp1}
\eeq
with $\quad n_i, n_i'=0,1,2,3,\ldots$ and we have introduced a set
of (yet unknown) functions $Q_n(x,y,z)$ depending on the three variables
$x=q^{-2n_1'}$, $y=q^{-2n_2'}$ and
$z=q^{-2n_3'}$. We notice that the formula (\ref{rp1}) contain
two conservations laws
\beq\label{con-law}
n_1+n_2=n_1'+n_2',\qquad n_2+n_3=n_2'+n_3',
\eeq
which are similar to the conservation law
of the 6-vertex model in two dimensions.

The specific $q$-dependent factor in \eqref{rp1}
has
been chosen to ensure that the functions $Q_n(x,y,z)$ are polynomials
in $x,y,z$ with coefficients which are themselves polynomials in the
variable $q$. They are completely determined by initial conditions
\beq\label{recur2}
Q_0(x,y,z)\equiv 1\,,\qquad \forall\ x,y,z=1,q^{-2},q^{-4},q^{-6}\ldots\,
\eeq
and the following recurrence
relation,
\beq
Q_{n+1}(x,y,z)=(x-1)\,(z-1)\,Q_n(x\,q^2,y,z\,q^2)
+x\,z\,(y-1)\,q^{2n}\,Q_n(x,y\,q^2,z)\,.\label{recur1}
\eeq
First two nontrivial polynomials  read
\beq\label{rp23}
\begin{array}{rcl}
Q_1(x,y,z)&=&1-(x+z)+x\,y\,z\,,\\[.3cm]
Q_2(x,y,z)&=&(1-x)\,(1-x\,q^2)\,(1-z)\,(1-z\,q^2)
-x^2\,z^2\,q^{4}\,(1-y^2)-\\[.3cm]
&&-x\,z\,q^2\,(1+q^2)\,(1-y)\,(1-x-z)\,.
\end{array}
\eeq
One can solve \eqref{recur1} with the
initial condition \eqref{recur2}
and derive the explicit formula valid for all values of $n$
\beq
Q_n(x,y,z)=(-x)^n q^{n(n-1)}\phis
(q^{-2n},\frac{q^{2-2n}}{xy};\frac{q^{2-2n}}{x};q^2,yz\,q^{2n})\label{rp4}
\eeq
where $\phis$ is the regularized terminating basic
hypergeometric series introduced in  (\ref{phi-reg}). This formula
works for any values $x,y,z=1,q^{-2},q^{-4},\ldots$.

Using (\ref{rp4}) one can rewrite (\ref{rp1}) in a more transparent form
convenient for further calculations
\beq
\begin{array}{rcl}
R_{\,n_1,\,n_2,\,n_3}^{{\,n'_1,\,n'_2,\,n'_3}^{\phantom{|}}}&=&
\delta_{n_1+n_2,n_1'+n_2'}\>\delta_{n_2+n_3,n_2'+n_3'}\,
q^{n_2(n_2+1)-(n_2-n_1')(n_2-n_3')}
\\[.5cm]
&&\ds\qquad\qquad\times
\sum_{r=0}^{n_2}\frac{(q^{-2n_1'};q^2)_{n_2-r}}{(q^2;q^2)_{n_2-r}}
\frac{(q^{2+2n_1};q^2)_r}{(q^2;q^2)_r}q^{-2r(n_3+n_1'+1)}\label{rp7}\, .
\end{array}
\eeq
As shown in \cite{MBS13} all nonzero matrix elements in (\ref{rp7}) are
positive for $0<q<1$.
The $R$-matrix
(\ref{rp7}) possesses the following symmetries
\beq
R_{\,n_1,\,n_2,\,n_3}^{\,n'_1,\,n'_2,\,n'_3}=
R_{\,n_3,\,n_2,\,n_1}^{\,n'_3,\,n'_2,\,n'_1},\quad
R_{\,n_2,\,n_1,\,n'_3}^{\,n'_2,\,n'_1,\,n_3}=
q^{n_2-n_1-n_3^2+{n'_3}^2}\frac{(q^2;q^2)_{n_3}}{(q^2;q^2)_{n'_3}}
R_{\,n_1,\,n_2,\,n_3}^{\,n'_1,\,n'_2,\,n'_3}
\label{rp7a}
\eeq
and solves the tetrahedron equation \cite{MBS13}
\beq\label{TE-op}
\Ru_{123}\,\Ru_{145}\,\Ru_{246}\,\Ru_{356}=
\Ru_{356}\,\Ru_{246}\,\Ru_{145}\,\Ru_{123}\,.
\eeq
It involves
operators acting in six Fock spaces, where $\Ru_{ijk}$ acts
non-trivially in the $i$-th, $j$-th and $k$-th spaces, but acts
as the identity in other three spaces.
In matrix form Eq.\eqref{TE-op} reads
\beq
\begin{array}{l}
{\ds\sum_{\stackrel[n'_4,\,n'_5,\,n'_6]{n'_1,\,n'_2,\,n'_3}{}}}
R^{n'_1\, n'_2\, n'_3}_{{n_1\, n_2\, n_3}^{\phantom{|}}}\
R^{n''_1\,n'_4\,n'_5}_{{n'_1\,n_4\,n_5}^{\phantom{|}}}\
R^{n''_2\,n''_4\,n'_6}_{{n'_2\,n'_4\,n_6}^{\phantom{|}}}\
R^{n''_3\,n''_5\,n''_6}_{{n'_3\,n'_5\,n'_6}^{\phantom{|}}}=\\[.8cm]
\qquad\qquad\qquad={\ds\sum_{\stackrel[n'_4,\,n'_5,\,n'_6]{n'_1,\,n'_2,\,n'_3}{}}}
R^{n'_3\,n'_5\,n'_6}_{{n_3\,n_5\,n_6}^{\phantom{|}}}\
R^{n'_2\,n'_4\,n''_6}_{{n_2\,n_4\,n'_6}^{\phantom{|}}}\
R^{n'_1\,n''_4\,n''_5}_{{n_1\,n'_4\,n'_5}^{\phantom{|}}}\
R^{n''_1\,n''_2\,n''_3}_{{n'_1\,n'_2\,n'_3}^{\phantom{|}}}\,.
\end{array}\label{rp8}
\eeq

Let $\lambda_i,\mu_i$, \ $i=1,2,\ldots,6$, be positive real numbers.
Using the conservation laws  it is easy to check that
if $\Ru_{ijk}$ satisfies \eqref{TE-op}, then so does the ``dressed''
$R$-matrix
\beq\label{R-new}
\Ru'_{ijk}=
\left(\frac{\mu_k}{\lambda_i}\right)^{\bosn_j} \Ru_{ijk}^{}
\left(\frac{\lambda_j}{\lambda_k}\right)^{\bosn_i}
\left(\frac{\mu_i}{\mu_j}\right)^{\bosn_k}\;,
\eeq
where the indices $(i,j,k)$ take four sets of values appearing in
\eqref{TE-op}. Note that the twelve parameters $\lambda_i,\mu_i$ enter the four
equations \eqref{R-new} only via eight independent ratios, so these
equations define a solution of \eqref{TE-op} containing
eight continuous parameters.
These new degrees of freedom  allow to
define a  non-trivial family of commuting layer-to-layer transfer
matrices.

In addition to \eqref{R-new} the tetrahedron equation
is, obviously, invariant under diagonal similarity transformations
\begin{equation}
\Ru'_{ijk}= c_i^{\bosn_i}\,c_j^{\bosn_j}\,c_k^{\bosn_k}\,\Ru_{ijk}\,
c_i^{-\bosn_i}\,c_j^{-\bosn_j}\,c_k^{-\bosn_k}.
\label{rp10}
\end{equation}
where $c_1,c_2,\ldots,c_6$ are arbitrary positive constants.

\nsection{The 2-layer projection and a composite $R$-matrix}

It is well known that any edge-spin model on the cubic lattice can be
viewed as a two-dimensional model on the square lattice with
an enlarged space of states for the edge spins
(see \cite{Bazhanov:2005as} for additional explanations).

Here we are going to exploit only the simplest 2-layer case. Consider
two vertices in the front-to-back direction as shown in Figure~1
where we also assume the
periodic boundary condition in the front-to-back direction.

\begin{figure}[ht]
\begin{picture}(200,120)(-100,-20)
{\thicklines
\put(80,25){\vector(1,0){50}}\put(120,65){\vector(1,0){50}}
\put(105,0){\vector(0,1){50}}
\put(145,40){\vector(0,1){50}}
\put(89,9){\vector(1,1){78}}}
\put(164,74){\footnotesize$k_1^{}$}\put(123,35){\footnotesize$k_2^{}$}
\put(80,4){\footnotesize$k_1^{}$}
\put(70,22){\footnotesize$i_1^{}$}\put(130,18){\footnotesize$i_1'$}
\put(107,-2){\footnotesize$j_1$}\put(95,50){\footnotesize$j_1'$}
\put(110,62){\footnotesize$i_2^{}$}\put(170,58){\footnotesize$i_2'$}
\put(147,38){\footnotesize$j_2^{}$}\put(135,90){\footnotesize$j_2'$}
\end{picture}
\caption{A front-to-back line of the cubic lattice}\label{rmat-fig}
\end{figure}
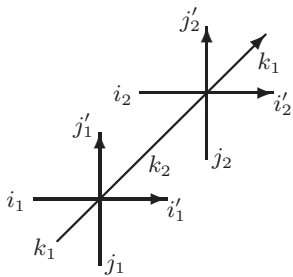
For further convenience we associate indices $\{j_1,j_2\}$, $\{j_1',j_2'\}$
with the first direction and $\{i_1,i_2\}$, $\{i_1',i_2'\}$ with the second direction.

Let us define a composite $R$-matrix
\beq
\mathbb{S}_{\boldsymbol{i}\,\,\boldsymbol{j}}^{\boldsymbol{i}'
  \boldsymbol{j}'}(w)=
\sum_{k_1,k_2}
{\Ru'\,}_{j_1^{},i_1^{},k_1^{}}^{j_1',i_1',k_{2}^{}}\,
{\widetilde\Ru'\,}_{j_2^{},i_2^{},k_2^{}}^{j_2',i_2',k_{1}^{}}.
\label{compR}
\eeq
The ``dressed'' $R$-matrices  $\Ru'$ and $\widetilde\Ru'$ used in (\ref{compR})
are derived from  $\Ru$ by combining both transformations (\ref{R-new}-\ref{rp10})
with different sets of fields in the first
and second directions, i.e. $\{c_i,\lambda_i,\mu_i\}$ and
$\{\widetilde c_i,\widetilde \lambda_i,\widetilde \mu_i\}$, $i=1,2$,
but with the same fields $\{c_3,\lambda_3,\mu_3\}$ in the front-to-back direction.
In the LHS of (\ref{compR}) we also introduced a new
``spectral'' parameter $w$ which is a special combination of fields  explicitly given below.

It follows from the conservation laws (\ref{con-law}) that we
can define two ``global'' conserved variables
\beq
I=i_1+i_2=i_1'+i_2',\quad J=j_1+j_2=j_1'+j_2'. \label{glob}
\eeq

Due to conservation laws (\ref{glob}) the $R$-matrix (\ref{compR})
acting in $(\mathcal{F}_q)^{\otimes2}\otimes(\mathcal{F}_q)^{\otimes2}$
decomposes into an infinite direct sum
\begin{equation}\label{decomp}
\mathbb{S}(w) \;=\; \mathop{\bigoplus}_{I,J=0}^\infty \ {\Rc}_{I,J}(w)
\end{equation}
 of the $U_q(\widehat{sl(2)})$
$R$-matrices with weights $I$ and $J$  ( or spins $I/2$ and $J/2$) \cite{MBS13}.
So fixing the values $I$ and $J$ in (\ref{glob}) we can derive
a general formula for matrix elements of such $R$-matrices.

Omitting some constant factors (depending on $I$ and $J$ and fields) one
can derive after simple calculations
\beq
\ds\big[\Rc_{I,J}(w)
\big]_{i_1,j_1}^{\ip,\jp}=\delta_{i_1+j_1,i_1'+j_1'}\,
\phi_h^{i_1}\phi_{v^{\phantom{1}}}^{j_1}
\psi_h^{i_1'-i_1}\psi_{v^{\phantom{1}}}^{j_1'-j_1}
\sum_{k_1,k_2}w^{k_1}{R\,}_{j_1^{},i_1^{},k_1^{}}^{j_1',i_1',k_{2}^{}}\,
{R\,}_{{J-j_1}^{},{I-i_1}^{},k_2^{}}^{{J-j_1'},{I-i_1'},k_{1}^{}},
\label{R-field}
\eeq
where
\beq
w=\frac{\mu_1\widetilde\mu_1}{\mu_2\widetilde\mu_2},
\quad \phi_h=\frac{\widetilde\lambda_1}{\lambda_1},
\quad \phi_v=\frac{\lambda_2}{\widetilde \lambda_2},\quad
\psi_h=\frac{\mu_2\widetilde c_2}{\mu_1 c_2},
\quad \psi_v=\frac{\widetilde c_1 \lambda_2}{c_1\widetilde \lambda_2}. \label{fields}
\eeq

The $R$-matrix (\ref{R-field}) satisfies the Yang-Baxter equation
\bea
&&\sum_{i_1',i_2',i_3'}
\big[\Rc_{I_1,I_2}(w)
\big]_{i_1,i_2}^{i_1',i_2'}
\big[\Rc_{I_1,I_3}'(w w')
\big]_{i_1',i_3}^{i_1'',i_3'}
\big[\Rc_{I_2,I_3}''(w')
\big]_{i_2',i_3'}^{i_2'',i_3''}=\nonumber\\
&&\sum_{i_1',i_2',i_3'}
\big[\Rc_{I_2,I_3}''(w')
\big]_{i_2,i_3}^{i_2',i_3'}
\big[\Rc_{I_1,I_3}'(w w')
\big]_{i_1,i_3'}^{i_1',i_3''}
\big[\Rc_{I_1,I_2}(w)
\big]_{i_1',i_2'}^{i_1'',i_2''},\label{YBEI3}
\eea
where $\Rc'$ and $\Rc''$ depend on different sets of fields
$\{\phi_h',\phi_v',\psi_h',\psi_v'\}$
and $\{\phi_h'',\phi_v'',\psi_h'',\psi_v''\}$.
These fields are not independent but satisfy the following constraints
\beq
\phi_v=\phi_v',\quad \phi_h'=\phi_h'',\quad \phi_v''=\phi_h^{-1},
\quad \psi_v''=\frac{\psi_h\psi_v'\psi_h''}{\phi_h\psi_v\psi_h'},\label{YBEfields}
\eeq
which easily follow from the conservation laws for the $R$-matrices
(\ref{R-field}) entering (\ref{YBEI3}) similar to the $6$-vertex model.
Equation (\ref{YBEI3}) acts in the tensor
product of three representation spaces with weights $I_1$, $I_2$, $I_3$ and is an immediate
consequence of the Yang-Baxter equation for the $q$-oscillator $R$-matrix
$\mathbb{S}(w)$ \cite{MBS13}.

All $\psi$'s fields  are simple gauge transformations of the $R$-matrix and do not affect
the spectrum of the transfer-matrix. Due to the conservation law in (\ref{R-field})
the transfer-matrix splits into a tensor sum of
blocks with equal sums of indices in the vertical direction.
Similar to the $6$-vertex model the vertical field $\phi_v$ will contribute the same
factor in each block and can be set to $1$ since it doesn't affect the spectrum.
As a result we get the following Yang-Baxter equation
\beq
\Rc_{I_1,I_2}(w;1)\Rc_{I_1,I_3}(w w';\phi_h)\Rc_{I_2,I_3}(w';\phi_h)=
\Rc_{I_2,I_3}(w';\phi_h)\Rc_{I_1,I_3}(w w';\phi_h)\Rc_{I_1,I_2}(w;1) \label{YBEred}
\eeq
where we explicitly showed a dependence on the horizontal field $\phi_h$.
As a consequence of (\ref{YBEred}) two transfer-matrices with the same  $\phi_h$
will commute.
From now on we shall assume that the $R$-matrix and the corresponding
transfer-matrix depend on the horizontal field.

Some extra care should be taken while calculating the sum in (\ref{R-field}). The summation
goes over all non-negative $k_1$, $k_2$
satisfying the condition
\beq
i_1+k_1=i_1'+k_2.\label{consind}
\eeq
For the case $i_1\geq i_1'$ we can exclude the index $k_2$ and safely sum over
$k_1$ from $0$ to $\infty$. However, for $i_1<i_1'$ the lower limit for $k_1$
should be $i_1'-i_1$. However, in this case
all contributions to the sum in (\ref{R-field}) from the values $0\leq k_1\leq i_1'-i_1-1$
are exactly zero. This happens because
the second $R$-matrix
in (\ref{R-field})
becomes zero
\beq
{R\,}_{{J-j_1}^{},{I-i_1}^{},k_2^{}}^{{J-j_1'},{I-i_1'},k_{1}^{}}=0 \label{zeroR}
\eeq
for $i_1<i_1'$ and $-(i_1'-i_1)\leq k_2\leq -1$.
So we can safely sum over   $k_1$  from $0$ to $\infty$ for all cases.
The property (\ref{zeroR}) cannot be
immediately seen from the definitions (\ref{rp1}) and (\ref{rp4}) and proved in Appendix A.

Now let us apply the second transformation from (\ref{rp7a}) to (\ref{R-field}). After
simple calculations we get the following symmetry of the matrix elements of the $R$-matrix
\beq
\big[{\Rc}_{J,I}(w;\{\phi\})
\big]_{j,i}^{j',i'}=q^{I-J}w^{i-i'}\big[\Rc_{I,J}(w;\{\phi^t\})
\big]_{i,j}^{i',j'},\label{symm1}
\eeq
where we introduced two sets of fields $\{\phi\}\equiv\{\phi_h,\phi_v,\psi_h,\psi_v\}$ and
$\{\phi^t\}\equiv\{\phi_v,\phi_h,\psi_v,\psi_h\}$.

We see that the $R$-matrix defined by (\ref{R-field}) is not completely symmetric with respect to
permutation of representation spaces with weights $I$ and $J$. We shall repair it below
by multiplying the $R$-matrix with the appropriate gauge and scalar factors.

Note that when we substitute the expression (\ref{rp7}) for the $R$-matrix into (\ref{R-field})
one can  see that the sum over $k_1$ converges provided
\beq
w < q^{I+J}.  \label{sum1}
\eeq
However, the resulting expression will be a rational function in $w$ which can be analytically
continued for any values of $w$.

Now let us remind that all nonzero elements of the 3D $R$-matrix (\ref{rp7})
are positive for $0<q<1$. Then nonzero matrix elements of the composite $R$-matrix
(\ref{R-field}) are also positive provided that field variables $\phi_h$,
$\phi_v$, $\psi_h$ and $\psi_v$ are positive and condition (\ref{sum1}) is satisfied.
Indeed, in this case the LHS of (\ref{R-field}) is given by a convergent series with positive
terms.

For further convenience
let us define new variables $\lam$ and $\phi$ and choose parameters in (\ref{R-field}) as
\beq
w=\lam^2,\quad \phi_h=\phi^2,\quad \phi_v=1,\quad \psi_h=1,\quad \psi_v=\lam.\label{sum2}
\eeq
With such a choice of fields
we define a properly normalized $R$-matrix by the following expression
\beq
\big[R_{I,J}(\lam)
\big]_{i,j}^{i',j'}=\sigma_{I,J}(\lam)q^I\,\big[\Rc_{I,J}(\lam)
\big]_{i,j}^{i',j'}\label{proper}
\eeq
where $\sigma_{I,J}(\lam)$ is symmetric in $I,J$ and defined by the following expression
\beq
\sigma_{I,J}(\lam)=(-1)^{m(I,J)}
q^{\frac{1}{2}I J-\frac{1}{2}m(I,J)}\lam^{-m(I,J)}(\lam^2q^{-I-J};q^2)_{m(I,J)+1},
\label{constf}
\eeq
where $m(i,j)=\min(i,j)$.

Finally, substituting  (\ref{rp7}) into (\ref{R-field})
we arrive at the following explicit formula
\beq\label{RIJ-def}
\begin{array}{l}
\ds\big[R_{I,J}(\lam;\phi)
\big]_{i,j}^{i',j'}\,=\,\ds\delta_{i+j,i'+j'}\,(-1)^{m(I,J)}\phi^{2i-I}\lam^{i-i'-m(I,J)}\,
\frac{q^{i^2+(I-i)(J-j')-i'(i'-j)+2I+\frac{1}{2}I J-\frac{1}{2}m(I,J)}}
{(q^2;q^2)_{i}\,(q^2;q^2)_{I-i}}
\\[.6cm]
\ds\qquad\times\  \
(\lam^2 q^{-I-J};q^2)_{m(I,J)+1}\
\sum_{k=0}^{i}\sum_{l=0}^{I-i}
\frac{(-1)^{k+l}\,q^{2k(i'-j)-2l(J-I-j+i)}}
     {q^{k(k+1)+l(l+1)}\ (1-\lam^2q^{\,I-J-2k-2l})^{\phantom{A^A}} }\\[.8cm]
\ds\qquad\times\ \frac{(q^{-2i},q^{2+2j};q^2)_k\
(q^{-2j'};q^2)_{i-k}}{(q^2;q^2)_k}
\frac{
(q^{-2(I-i)},q^{2(1+J-j)};q^2)_l\ (q^{-2(J-j')};q^2)_{I-i-l}}{(q^2;q^2)_l}
\end{array}
\eeq
with
\beq
0\le i,i'\le I,\qquad
0\le j,j'\le J\,.
\eeq
One can show that with such a normalization the matrix elements of
the $R$-matrix are the polynomials in $\lam$ and $\lam^{-1}$ of the  degree
$\leq m(I,J)$. We shall justify the
choice of normalization (\ref{constf}) in the next section.

\nsection{Properties of the higher spin $R$-matrix}

In this section we shall analyze  the formula (\ref{RIJ-def}) and derive
a remarkably simple formula for the $R$-matrix $R_{I,J}(\lambda)$ in the form of a single sum.

First let us notice the formula (\ref{RIJ-def}) can be naturally extended to
any  values $J\in\mathbb{C}$. This corresponds to the case when the second space
becomes an infinite-dimensional Verma module $V_J^+$ with indices $j, j'$ running
form $0$ to $\infty$. In this case  we will choose
\beq
m(I,J)=I, \quad I\in\mathbb{Z}_+, \quad J\in\mathbb{C}.\label{prop1}
\eeq

As an example consider the case $I=1$. The double sum in (\ref{RIJ-def})
contains only two nontrivial terms and we obtain

\beq
[R_{(1,J)}(\lambda;\phi)]_{i,j}^{i',j'}=
\left(\begin{array}{ll}
\delta_{j,j'}\phi^{-1}[\lambda q^{\frac{1+J}{2}-j'}]&
\delta_{j,j'+1}\phi^{-1}[q^{J-j'}]q^{j'-\frac{J-1}{2}}\\
& \\
\delta_{j+1,j'}\phi\,[q^{j'}]q^{\frac{J+1}{2}-j'}&\delta_{j,j'}\phi\,
[\lambda q^{\frac{1-J}{2}+j'}]\\
\end{array}\right)_{i+1,i'+1},\label{prop2}
\eeq
where $i,i'=0,1$.

As usual we can define the $L$-operator
 acting in the tensor product $\mathbb{C}^2\otimes V^+_J$ as a two-by-two matrix
with matrix elements coinciding with the matrix elements of $R_{1,J}(\lambda)$.
To make a connection to the standard XXZ $L$-operator we
introduce a rescaled spectral parameter $\mu=\lambda q^{1/2}$, set $\phi=1$
and apply a simple similarity transformation $D=\mbox{diag}(1,\lambda^{-1})$
in $\mathbb{C}^2$. Then we obtain
\beq
L(\mu)=
\left(\begin{array}{ll}\ds
\mu q^{H/2}-\mu^{-1}q^{-H/2}&
\mu [q]\,F q^{-H/2}\\
& \\
\mu^{-1}[q]\,q^{H/2}E &\mu q^{-H/2}-\mu^{-1}q^{H/2}\\
\end{array}\right)\label{prop4}
\eeq
where $E$, $F$ and $H$ are the generators of the quantum algebra $U_q(sl(2))$
with the action (\ref{conv5}).

If $J\in\mathbb{Z}_+$, the module $V_J^+$ becomes reducible and first $J+1$ vectors
$\{v_0,\ldots, v_J\}$ form a basis of a $(J+1)$-dimensional representation space $V_J$.

Now let us assume that $J$ is not a positive integer and perform a resummation
 in (\ref{RIJ-def}) by introducing a new variable
$s=k+l$. Then we can rewrite the sum in (\ref{RIJ-def}) in the form of a pole expansion
in $\lam^2$ at the points $\lam^2=q^{J-I+2s}$, $s=0,\ldots,I$.
Surprisingly the corresponding residues can be expressed in terms of terminating balanced
$\phifs$ series.
Applying Sears' transformation for terminated balanced $\phifs$
(\ref{app2}) from Appendix B
we can rewrite (\ref{RIJ-def}) in the following form
\beq\label{Rphi}
\begin{split}
&[R_{I,J}(\lam;\phi)]_{i,j}^{i',j'}=\delta_{i+j,i'+j'}(-1)^{i+I}
\phi^{2i-I}\lam^{i-i'-I}q^{i(i+I-2J-1)+(I-i)(J-j)+i'(I+j')+I(3+J)/2}
\times
\\
&
\times{\ds \frac{(q^{-2J};q^2)_j}{(q^{-2J};q^2)_{j'}}
\frac{(\lam^2 q^{-I-J};q^2)_{I+1}}{(q^2;q^2)_{i}}
\sum_{s=0}^{I}\,\frac{(-1)^s}{1-\lam^2q^{I-J-2s}}}
\frac{q^{s(s-1)-2is}}{(q^2;q^2)_s(q^2;q^2)_{I-s}}c_{i,j}^{i',j'}(I,J;s)
\end{split}
\eeq
with coefficients $c_{i,j}^{i',j'}(I,J;s)$  given by
\beq
c_{i,j}^{i',j'}(I,J;s)=\frac{(q^{2(I-J-s)};q^2)_{j'-i}(q^{-2(s+J)};q^2)_I}
{(q^{-2(s+J)};q^2)_{i+j}}{\phif\left(\left.\begin{array}{l}
q^{-2i};q^{-2i'},q^{-2s},q^{2(1+J-I+s)}\\
q^{-2I},q^{2(1+j'-i)},q^{2(1+J-i-j)}\end{array}\right|q^2,q^2\right)},\label{prop8}
\eeq
where we used our definition of regularized terminating hypergeometric series (\ref{phi-reg}).

We note that the only problem for integer $J$ comes from possible poles in
(\ref{prop8}) for $0\leq s<i+j-J$. However, one can show that for all such values
of $s$ there is exactly a matching zero coming from $\phif$. So if we define coefficients
$c_{i,j}^{i',j'}(I,J;s)$ for integer $J$ as a limiting value from complex $J$, then
the representation
(\ref{Rphi}) for matrix elements works for any integer $I$ and $J$ provided that $I\leq J$.

Now let us notice the  sum in (\ref{Rphi}) can be represented as a ratio
\beq
\frac{R_I(\lambda^2)}{(\lam^2 q^{-I-J};q^2)_{I+1}} \label{prop9}
\eeq
where $R_I(\lambda^2)$ is a polynomial of the degree $I$ in $\lambda^2$.
Such a polynomial can be reconstructed using a Lagrange interpolation formula.
Applying this formula for any polynomial $P_n(x)$ of the degree $n$ one can easily show that
\beq
x^{n+1}(x^{-1};q)_{n+1} \sum_{i=0}^n(-1)^i\frac{1}{x-q^i}
\frac{ q^{i(i+1)/2-ni}}{(q;q)_i(q;q)_{n-i}}P_n(q^i)=P_n(x).\label{prop9a}
\eeq
This allows us to perform a summation over $s$ in (\ref{Rphi}) and obtain the main result
of this paper
\beq
[R_{I,J}(\lam;\phi)]_{i,j}^{i',j'}=\delta_{i+j,i'+j'}\,
\phi^{2i-I}
a_{ij}^{i'j'}(\lambda)
\phif\left(\left.\begin{array}{l}
q^{-2i};q^{-2i'},\lambda^{-2} q^{J-I},\lambda^{2}q^{2+J-I}\\
\phantom{aa}q^{-2I},q^{2(1+j-i')},q^{2(1+J-i-j)}\end{array}\right|q^2,q^2\right)\label{prop10}
\eeq
where
\beq
a_{ij}^{i'j'}(\lambda)=(-1)^{i}\,\frac{
 q^{i(i-J-1)+(I-i)(J-j)+i'(I+j')}}{\lambda^{i+i'-m(I,J)}\,q^{\frac{1}{2}IJ-\frac{1}{2}m(I,J)}}
 \frac{(q^{-2J};q^2)_j}{(q^{-2J};q^2)_{j'}}
\frac{(\lambda^{-2}q^{I-J};q^2)_{j-i'}(\lambda^{-2}q^{-I-J};q^2)_{m(I,J)}}
{(q^2;q^2)_i(\lambda^{-2}q^{-I-J};q^2)_{i+j}}.\label{prop11}
\eeq

Let us make a few important remarks regarding  (\ref{prop10}). First, it is easy to see that
the main ingredient of  (\ref{prop10}) comes from the coefficients $c_{i,j}^{i',j'}(I,J;s)$
with $q^{2s}$ replaced by $\lambda^2q^{I-J}$. Assuming that $\lambda$ is generic we no longer have
poles coming from the pre-factor in (\ref{prop8}).
So (\ref{prop10}) is well defined for integer values of $J$ as well.

 Second,  we derived the formula (\ref{prop10}) assuming that
$m(I,J)=I$, i.e. $I\leq J$ for $I,J\in\mathbb{Z}_+$. However, as we already know from
(\ref{symm1}),  the formula for the matrix elements should have a certain symmetry
with respect to interchanging $I$ and $J$. Namely, we should have
\beq
\mathcal{P}_{12}R_{I,J}(\lam;1)\mathcal{P}_{12}=R_{J,I}(\lam;1),\label{prop12}
\eeq
where $\mathcal{P}_{12}$ is the permutation operator.
This symmetry of the $R$-matrix immediately follows from the symmetry transformation (\ref{app4})
for $\phif$ basic hypergeometric series derived in the Appendix B. It also
allows us to restore all factors $m(I,J)$ correctly which were originally set to $I$.

The next comment concerns the normalization of the $R$-matrix $R_{I,J}(\lambda;\phi)$.
One can easily derive from (\ref{prop10}) that
\beq
[R_{I,J}(\lam;\phi)]_{0,0}^{0,0}=\phi^{-I}q^{\frac{1}{2}IJ+\frac{1}{2}m(I,J)}\lambda^{m(I,J)}
(\lambda^{-2}q^{-I-J};q^2)_{m(I,J)}.\label{prop13}
\ee
We  see that up to an overall normalization factor  matrix elements
of the $R$-matrix are rational functions
in $q^I$ and $q^J$ and polynomials in $\lambda$ and $\lambda^{-1}$ of the degree
determined by indices $i,j,i',j'$.

Now we can consider three different cases. If both $I, J\in\mathbb{Z}_+$, all matrix elements are
polynomials in $\lambda$ and $\lambda^{-1}$ of the degree $d\leq m(I,J)$.
This is also true in the case when
only $I$ (or $J$)
is a positive integer and  we set $m(I,J)=I$ (or $m(I,J)=J$).

However, the formula (\ref{prop10}) works even in the case when the $R$-matrix acts
in the tensor product of two infinite-dimensional Verma modules $V^+_I\otimes V^+_J$,
$I,J\in\mathbb{C}$.
In this case we can choose a different normalization of the $R$-matrix, say,
\beq
[R_{I,J}(\lam;1)]_{0,0}^{0,0}=1.\label{prop14}
\eeq

To confirm the last statement we could define the $R$-matrix $R_{I,J}(\lam;1)$
in a zero field as a solution
of the  Yang-Baxter equation acting in $\mathbb{C}^2\otimes V_I^+\otimes V_J^+$ ,
$I,J\in\mathbb{C}$
\beq
L_{1,I}(\mu)L_{1,J}(\lambda\mu)R_{I,J}(\lambda;1)=
R_{I,J}(\lambda;1)L_{1,J}(\lambda\mu)L_{1,I}(\mu),
\label{prop15}
\eeq
where the $L$-operators $L_{1,I}(\lambda)$ and $L_{1,J}(\mu)$ are defined as in
(\ref{prop4}). Substituting
\beq
[R_{I,J}(\lambda;1)]_{i,j}^{i',j'}=\delta_{i+j,i'+j'}S_{i,j}^{i',j'}\label{prop16a}
\eeq
into (\ref{prop15}) we obtain the system of  three
linearly independent recursions. These recurrence relations are given in
Appendix C, (\ref{Crecur1}-\ref{Crecur3}).
Up to a normalization $S_{0,0}^{0,0}$ they have a unique solution which
coincides with (\ref{prop10}).
However, it is highly nontrivial to find a solution of (\ref{Crecur1}-\ref{Crecur3})
in terms of basic hypergeometric series.
Our derivation of (\ref{prop10}) is based on the 3D approach
where it appears very naturally.

Now let us return to the case when $I,J\in\mathbb{Z}_+$. Then the $R$-matrix
$R_{I,J}(\lambda;\phi)$ has another symmetry
\beq
[R_{I,J}(\lambda;\phi)]_{i,j}^{i',j'}=
[R_{I,J}(\lambda;\phi^{-1})]_{I-i,J-j}^{I-i',J-j'}\label{prop16}
\eeq
which will be used to define the second $Q$-operator in the next section.

Note that in the case $I=J=1$ (\ref{prop16}) is equivalent to the invariance of the $R$-matrix
under the conjugation by the operator $\mathcal{R}=\sigma_x\,\otimes\,\sigma_x$ and
the transformation $\phi\to\phi^{-1}$.
The proof of the relation (\ref{prop16}) is reduced to applying the  Sears transformation
(\ref{app3}) to (\ref{prop10}).

Additionally the $R$-matrix possesses the following
symmetry  under a simultaneous transformation $\lambda\to\lambda^{-1}$ and $q\to q^{-1}$:
\beq
R_{I,J}(\lambda^{-1};\phi)|_{q\to q^{-1}}=
(-1)^{m(I,J)}D_1\otimes D_2 \, R_{I,J}(\lambda;\phi) \, D_1^{-1}\otimes D_2^{-1},\label{prop17}
\eeq
where
diagonal matrices $D_1$ and $D_2$ acting in $V_I$ and $V_J$ are
\beq
[D_1]_{i,i'}=\delta_{i,i'}q^{-i(i-1)},\quad i=0,\ldots,I,\quad [D_2]_{j,j'}=\delta_{j,j'}
q^{-j(j-1)+j(J-I)},\quad j=0,\ldots,J.\label{prop18}
\eeq
This symmetry can be proved by observing that the defining relations for the $R$-matrix
(\ref{Crecur1}-\ref{Crecur3})
are invariant under combined transformations from (\ref{prop17}).

In the case $I=J$ and $\lambda=\phi=1$ the $R$-matrix reduces to the permutation operator
\beq
R_{I,I}(1;1)=q^{-\frac{1}{2}I(I+1)}(q^2;q^2)_I\,\mathcal{P}_{12}\label{prop19}
\eeq
which can be proved directly from the formula (\ref{prop10}).

Finally, one can calculate the expansions of (\ref{prop10}) near the points $\lambda=0$ and
$\lambda=\infty$. In the leading order we get
\beq
[R_{I,J}(\lambda;\phi)]_{i,j}^{i',j'}=\delta_{i,i'}\delta_{j,j'}(-\lambda q^{1/2})^{-m(I,J)}
\phi^{2i-I} q^{-\frac{1}{2}(I-2i)(J-2j)}\left(1+O(\lambda)\right)
\quad \mbox{at}\quad \lambda\to0\label{prop20}
\eeq
and
\beq
[R_{I,J}(\lambda;\phi)]_{i,j}^{i',j'}=\delta_{i,i'}\delta_{j,j'}(\lambda q^{1/2})^{m(I,J)}
\phi^{2i-I} q^{\frac{1}{2}(I-2i)(J-2j)}\left(1+O(\lambda^{-1})\right)
\quad \mbox{at}\quad \lambda\to\infty.\label{prop21}
\eeq

We could calculate next order corrections in (\ref{prop21})
and compare it with Kirillov and Reshetikhin expansion at $\lambda\to\infty$
 in \cite{KR87a}. We leave this exercise to the reader.

\nsection{Q-operators}

The theory of the $Q$-operators related to the affine algebra $U_q(\widehat{sl(2)})$ has been
developed in \cite{BLZ97a,BLZ99a}. Two $Q$-operators ${\bf{Q}}_\pm(\lambda)$ appear
as traces of special
monodromy matrices over infinite-dimensional representations of the $q$-oscillator algebra
introduced in Section 2. Similar to the usual transfer-matrices these monodromy matrices
are derived from the tensor product of the local $L$-operators.
In this section we construct  these local $L$-operators acting in the
$(I+1)$-dimensional highest weight module $V_I$. We also show that
this construction can be naturally generalized to the case of
infinite-dimensional Verma module ${V}^+_I$ with $I\in\mathbb{C}$.

Lets us first assume that $I\in\mathbb{Z}_+$. Then we can introduce the transfer matrix
${\bf\widehat{T}}_{J,I}(\lambda;\phi)$ associated with the infinite-dimensional
Verma module, ${V}_J^+$,
$J\in\mathbb{C}$ acting in the quantum space $ W=\stackrel[i=1]{M}{\otimes}V_I$ as
\beq
{\bf\widehat{T}}_{J,I}(\lambda;\phi)=\underset{{{V}^+_J}}{\mbox{ Tr}}
\big[\underbrace{R_{J,I}(\lambda;\phi)
\otimes...\otimes R_{J,I}(\lambda;\phi)}_{M\>\text{\small times}}\big].\label{qops1}
\eeq
where the trace is defined as in (\ref{conv7}).

We notice that usually the horizontal field  is introduced via a global twist in
the auxiliary space \cite{BLZ97a, BLZ99a}.  However, we prefer to use a local field  and
include it into the definition of the $R$-matrix. With such a definition the transfer-matrix
still commutes with the shift operator along the periodic chain. We are going to exploit
this fact in our construction of factorized $Q$-operators in the next publication.

Due to the conservation law in (\ref{prop10}) the transfer matrix (\ref{qops1})
has a block-diagonal form
\beq
\ds{\widehat{\bf T}}_{J,I}(\lambda;\phi)=
\ds\stackrel[l=0]{IM}{\ds\bigoplus}{\widehat{\bf T}}_{J,I}^{(l)}(\lambda;\phi),\label{qops2}
\eeq
where for each block ${\widehat{\bf T}}_{J,I}^{(l)}(\lambda;\phi)$ the sum of
in- and out- indices in the quantum space $W$ is fixed to $l$, i.e.
\beq
\sum_{k=1}^M i_k=\sum_{k=1}^M i'_k=l. \label{qops3}
\eeq
Let us call the subspace in the quantum space $W$ with a fixed $l$ as the $l$-th sector.

The direct sum expansion (\ref{qops2}) is also true for $I\in\mathbb{C}$,
when the quantum space becomes infinite-dimensional, $W=\stackrel[i=1]{M}{\otimes}V_I^+$.
In this case the sum in (\ref{qops2}) runs from zero to infinity, but all blocks with a fixed $l$
are still finite-dimensional.

Using asymptotics (\ref{prop20}-\ref{prop21})
one can easily calculate asymptotics of the ${\bf\widehat{T}}_J^{(I)}(\lambda;\phi)$
in each block with a fixed $l$
\beq
{\bf\widehat{T}}_{J,I}^{(l)}(\lambda;\phi)|_{\lambda\to0}=
(-\lambda q^{1/2})^{-I M}\phi^{-JM}\frac{q^{-\frac{1}{2}I J M+J l}}{1-\phi^{2M}q^{I M-2l}}
\left({\boldsymbol I}_l+O(\lambda)\right),\label{qops4a}
\eeq
\beq
{\bf\widehat{T}}_{J,I}^{(l)}(\lambda;\phi)|_{\lambda\to\infty}=(\lambda q^{1/2})^{I M}
\phi^{-JM}\frac{q^{\frac{1}{2}I J M-J l}}{1-\phi^{2M}q^{2l-I M}}
\left({\boldsymbol I}_l+O(\lambda^{-1})\right),\label{qops4}
\eeq
where ${\boldsymbol I}_l$ is the unit matrix of the dimension of the block.

When $J\in\mathbb{Z}_+$, the module $V_J^+$ becomes reducible as discussed in  Section 2.
The formula for the $R$-matrix (\ref{prop10}) is analytic in $q^J$ except the normalization factor
which contain the function $m(I,J)$. The transfer matrix
${\widehat{\bf T}}_{J,I}(\lambda;\phi)$ splits into two terms
\beq
{\widehat{\bf T}}_{J,I}(\lambda;\phi)={{\bf T}}_{J,I}(\lambda;\phi)+
{\widehat{\bf T}}_{-J-2,I}(\lambda;\phi), \label{qops5}
\eeq
where ${{\bf T}}_{J,I}(\lambda;\phi)$ is the transfer matrix defined similar to (\ref{qops1})
with the trace taken over the finite-dimensional representation $\pi_J$.

However, there is one subtlety related to our choice of  normalization (\ref{prop13}).
When $J$ is not integer, we assumed that $m(I,J)=I$ for $I\in\mathbb{Z}_+$ in our
definition of the transfer-matrix (\ref{qops1}). However,
when $J$ becomes a positive integer, we can expect in (\ref{qops5}) some extra factor in front
of ${{\bf T}}_{J,I}(\lambda;\phi)$ for the case $I>J$, since $m(I,J)=J$ in this case.
Indeed, a detailed analysis shows
that we need to slightly modify (\ref{qops5}) as follows
\beq
h_{I-J}(\lambda)^M {{\bf T}}_{J,I}(\lambda;\phi)=
{\widehat{\bf T}}_{J,I}(\lambda;\phi)-{\widehat{\bf T}}_{-J-2,I}(\lambda;\phi), \label{qops6}
\eeq
where
\beq
h_{I}(\lambda)=\prod_{k=0}^{I-1}\big[\lambda q^{\frac{I}{2}-k}\big]=
\begin{cases}
1,& \text{for $I\leq 0$,}\\
(\lambda q^{1/2})^{I}(\lambda^{-2}q^{-I};q^2)_{I},& \text{for $I>0$.}
\end{cases}
\label{qops7}
\eeq

Let us illustrate  equation (\ref{qops6}) with the case $I=M=1$.
Using formula (\ref{prop10}) we obtain after simple calculations
\beq
{\widehat{\bf T}}_{J,1}(\lambda;\phi)=
\begin{pmatrix} \frac{\ds\lambda \phi^{-J}q^{\frac{1+J}{2}}}{\ds1-\phi^2/q}-
\frac{\ds\lambda^{-1} \phi^{-J}q^{-\frac{1+J}{2}}}{\ds1-\phi^2 q} & 0\\
0 & \frac{\ds\lambda \phi^{-J}q^{\frac{1-J}{2}}}{\ds1-\phi^2q}-
\frac{\ds\lambda^{-1} \phi^{-J}q^{\frac{J-1}{2}}}{\ds1-\phi^2/q}
\end{pmatrix}\label{qops8}
\eeq

Then for any integer $J\geq1$ we get from (\ref{qops6})
\beq
{\bf T}_{J,1}(\lambda;\phi)=
\begin{pmatrix}
\phi^{-J}\ds\sum_{k=0}^J\phi^{2k}\big[\lambda q^{\frac{1+J}{2}-k}\big]
 & 0\\
0 & \phi^{-J}\ds\sum_{k=0}^J\phi^{2k}\big[\lambda q^{\frac{1-J}{2}+k}\big]
\end{pmatrix}\label{qops9}
\eeq
which can be verified by direct calculations using the explicit formula (\ref{prop10})
for the $R$-matrix $R_{J,I}(\lambda;\phi)$.

We also notice the following normalization of the transfer-matrix for $J=0$ and
$I\in\mathbb{Z}_+$
\beq
{\bf T}_{0,I}(\lambda;\phi)={\boldsymbol I},\label{qops10}
\eeq
which is an immediate consequence of (\ref{prop10}).

Now we turn to the construction of the ${Q}$-operators for any highest
weight representation in the quantum space. The main algebraic properties
of the ${Q}$-operators are encoded into the
{\it fundamental fusion relation} discovered in \cite{BLZ97a,BLZ99a}
\beq
\mbox{\bf Wr}(\phi){\bf\widehat{T}}_{J,I}(\lambda;\phi)=
\ds{\bf Q}^{(I)}_+(\lambda q^{-\frac{J+1}{2}})
{\bf Q}^{(I)}_-(\lambda q^{\frac{J+1}{2}}),\label{qops11}
\eeq
where the Wronskian $\mbox{\bf Wr}(\phi)$
does not depend on the spectral parameter $\lambda$.
The Wronskian is the diagonal operator and its eigenvalues are the same
in every $l$-th sector.
We notice that our
factorization relation (\ref{qops11})
is slightly different from the one used in \cite{BLZ97a,BLZ99a} due to
the fact that their $\lambda$ is, in fact, our $\lambda^{-1}$.

The relation (\ref{qops11}) has been derived in \cite{BLZ97a,BLZ99a}
irrespective of the choice of the quantum space using the universal $R$-matrix approach
\cite{Drinfeld1985}.
In principle, using the explicit expression of the $U_q(\widehat{sl(2)})$
universal $R$-matrix
\cite{TK92} one can construct the ${ Q}$-operators for any highest
weight representation. However, calculations for higher spins quickly become unbearable.
In our approach we derive the ${Q}$-operators for an
arbitrary weight $I$ based on the explicit construction of
the $R$-matrix (\ref{prop10}).

Although the eigenvalues of the transfer-matrix are polynomials in
$\lambda$ and $\lambda^{-1}$,
the eigenvalues of the $Q$-operators are not. In the twisted case $\phi\neq1$
they are polynomials multiplied by simple exponential factors
depending on the horizontal field.
Namely, set
\beq
\lambda=e^{iu}, \quad \phi=q^h \label{qops12a}
\eeq
and define two operators ${\bf A}^{(I)}_\pm(\lambda)$ as
\beq
{\bf Q}^{(I)}_\pm(\lambda)=e^{\pm i u h M}{\bf A}^{(I)}_\pm(\lambda)=
\lambda^{\pm h M}{\bf A}^{(I)}_\pm(\lambda), \label{qops12b}
\eeq
where $M$ is the size of the system.
Then the  eigenvalues of the operators ${\bf A}^{(I)}_\pm(\lambda)$ will be
polynomials in $\lambda$ and $\lambda^{-1}$ (see, for example, \cite{BM07} for detailed
explanations).

Then we can rewrite the relation (\ref{qops11}) as
\beq
\phi^{(J+1)M}\mbox{\bf Wr}(\phi){\bf\widehat{T}}_{J,I}(\lambda;\phi)=
\ds{\bf A}^{(I)}_+(\lambda q^{-\frac{J+1}{2}})
{\bf A}^{(I)}_-(\lambda q^{\frac{J+1}{2}}).\label{qops12c}
\eeq
Let us now make the following substitution into (\ref{qops12c})
\beq
\lam=\mu q^{-\frac{J+1}{2}}\label{qops12d}
\eeq
and consider the limit $J\to\infty$ with $\mu$ being fixed. In the RHS of
(\ref{qops12c}) we shall get
the operator ${\bf A}^{(I)}_-(\mu)$ pre-multiplied by a constant matrix
${\bf A}^{(I)}_+(\infty)$. We can always absorb this matrix into ${\bf A}^{(I)}_-(\mu)$ by
redefining it , so we assume that
\beq
{\bf A}^{(I)}_+(\infty)\sim {\boldsymbol I}\label{qops12}
\eeq
and check a consistency of (\ref{qops12}) later. Now taking the limit $J\to\infty$
and substituting
(\ref{qops12d}) in (\ref{prop10}) we define the local $L$-operator $A^{(I)}_-(\mu)$
\beq
[A^{(I)}_-(\mu)]_{n,i}^{n',i'}=\lim_{J\to\infty}\left\{[R_{J,I}(\mu q^{-\frac{J+1}{2}};\phi)]_{n,i}^{n',i'}U_{i,i'}^n(I,J)\right\}
\label{qops13}
\eeq
where
\beq
U_{i,i'}^n(I,J)=\phi^J(-1)^{II-i}
\lambda^{-i'} q^{(i-i')(j+j')+J(3i'-i)/2+(i'-i)/2}. \label{qops14}
\eeq

Let us notice that we introduced some additional factor (\ref{qops14}) into (\ref{qops13}).
It is needed to compensate  divergent contributions in $J$ and simplify the formula
for the $L$-operator. It is easy to prove the following identity
\beq
\sum_{k=1}^M (i_k-i'_k)(n_k+n'_k) =\biggl[\sum_{k=1}^M (i_k-i'_k)\biggr]^2\label{qops15}
\eeq
provided that $i_k+n_k=i'_k+n'_k$, $k=1,\ldots,M$. Therefore, the factor $U_{i,i'}^n(I,J)$
can only contribute a constant to each block with a fixed $l$ (see (\ref{qops3})).

After simple calculations we obtain the following result
\beq
\begin{split}
[A^{(I)}_-(\lambda)]_{n,i}^{n',i'}=&\delta_{i+n,i'+n'}\,{\phi^{2n}}\lambda^{i-I}
q^{i I+ii'+n(I-i-i')}\times \\
&\times\frac{(\lambda^2q^{1-I+2(i'-n)};q^2)_{I-i-i'}}
{(q^2;q^2)_i\,\,}\,
{\prescript{}{3}{\overline\phi}}_2\left(\left.\begin{array}{l}
q^{-2i};q^{-2i'},\lambda^2q^{1-I}\\
q^{-2I},q^{2(1+n-i')}\end{array}\right|q^2,q^2\right).
\end{split}\label{qops16}
\eeq

The formula (\ref{qops16}) defines a local $L$-operator $A^{(I)}_-(\lambda)$
acting in the tensor
product $\mathcal{F}_q\otimes V_I$. To define the corresponding
global ${Q}$-operator we take a tensor product of $M$ copies of (\ref{qops14}),
take a normalized trace (\ref{conv8}) over the Fock space $\mathcal{F}_q$
and multiply by the exponential factor from (\ref{qops12b})
\beq
{\bf Q}^{(I)}_-(\lambda)=\lambda^{-h M}\,
\underset{{\,\mathcal{F}_q}}{\widehat{\mbox{Tr}}}\,\{
\underbrace{
A^{(I)}_-(\lambda)\otimes\ldots\otimes A^{(I)}_-(\lambda)}_\text{$M$ times}\}.\label{qops17}
\eeq

We could use the same strategy and define the second ${Q}$-operator
${\bf Q}^{(I)}_+(\lambda)$
by taking
a different limit in (\ref{qops12c}). However, we prefer to use a different approach.
Let us remind that the operators ${\bf Q}^{(I)}_\pm(\lambda)$ are two linearly
independent solutions of the $TQ$-relation (which we prove later) with the transfer
matrix
\beq
{\bf{T}}_{1,I}(\lambda;\phi)=\underset{{{V}_1}}{\mbox{ Tr}}
\big[\underbrace{R_{1,I}(\lambda;\phi)
\otimes...\otimes R_{1,I}(\lambda;\phi)}_{M\>\text{\small times}}\big].\label{qops18}
\eeq
This transfer matrix has the following symmetry
\beq
{\bf{T}}_{1,I}(\lambda;\phi)_{i_1,\ldots,i_M}^{i'_1,\ldots,i'_M}=
{\bf{T}}_{1,I}(\lambda;\phi^{-1})_{I-i_1,\ldots,I-i_M}^{I-i'_1,\ldots,I-i'_M},
\label{qops19}
\eeq
which is a consequence of the symmetry (\ref{prop16}) of the $R$-matrix.

It  follows that we can define the second $L$-operator
\beq
[A^{(I)}_+(\lambda)]_{n,i}^{n',i'}=
[A^{(I)}_-(\lambda)]_{n,I-i}^{n',I-i'}|_{\phi\to\phi^{-1}}\label{qops20}
\eeq
and the second $Q$-operator ${\bf Q}^{(I)}_+(\lambda)$ by
\beq
{\bf Q}^{(I)}_+(\lambda)=
\lambda^{h M}\,\underset{{\,\mathcal{F}_q}}{\widehat{\mbox{Tr}}}\,\{
\underbrace{
A^{(I)}_+(\lambda)\otimes\ldots\otimes
A^{(I)}_+(\lambda)}_\text{$M$ times}\}.\label{qops21}
\eeq
Due to the symmetry (\ref{qops19}) ${\bf Q}^{(I)}_+(\lambda)$ will
satisfy the same $TQ$-relation.
Since $TQ$-relation has only two linear independent solutions and
${\bf Q}^{(I)}_\pm(\lambda)$
cannot mix\footnote{The operators ${\bf Q}^{(I)}_\pm(\lambda)$ satisfy different
quasi-periodicity conditions
under the shift $u\to u+\pi$ due to (\ref{qops12b})},
we have constructed the second
$Q$-operator.

In principle, (\ref{qops20}) completely determines matrix elements of
the $L$-operator
$A^{(I)}_+(\lambda)$. However, it has a big disadvantage, since it can
be applied only for integer
values of $I$. In fact, using the transformation (\ref{app5}) for
hypergeometric series
$\phit$ one can transform (\ref{qops20}) to the following neat form
\beq
\begin{split}
[A^{(I)}_+(\lambda)]_{n,i}^{n',i'}=
\delta_{i+n',i'+n}&\,{\phi^{-2n}}(-1)^{i+i'}\lambda^{-i}
q^{i(i+1)-i'(i'+1)+i'(I+i)+n(I-i-i')}\times \\
&\times\frac{(q^{2};q^2)_{n'}}
{(q^2;q^2)_n(q^2;q^2)_i\, }\,
{\prescript{}{3}{\overline\phi}}_2\left(\left.\begin{array}{l}
q^{-2i};q^{-2i'},\lambda^2q^{1-I}\\
q^{-2I},q^{2(1+n-i)}\end{array}\right|q^2,q^2\right).\label{qops22}
\end{split}
\eeq
Comparing (\ref{qops16}) and (\ref{qops22}) we observe that the  $L$-operator
$A^{(I)}_+(\lambda)$ coincides with $A^{(I)}_-(\lambda)^{t_2}$ up to a
simple diagonal transformation,
transformation $\phi\to\phi^{-1}$
and equivalence transformations in both auxiliary and quantum space.

Let us comment on the obtained results. First, it is clear that both $Q$-operators
${\bf Q}^{(I)}_\pm(\lambda)$
have the same
block-diagonal form as the transfer-matrix (\ref{qops18}) due to the presence of the
delta-functions in
(\ref{qops16}) and (\ref{qops22}). Second, ${\bf Q}^{(I)}_\pm(\lambda)$ commute with the
transfer-matrix and this is the consequence of the Yang-Baxter equation
for the $R$-matrix (\ref{prop10}).

Finally, it is clear  that the operator ${\bf A}^{(I)}_+(\lambda)$
is well defined even for $I\in\mathbb{C}$, since the dependence on $I$ is
analytic in (\ref{qops22}). The matrix elements of this operator are always polynomials
in $\lambda$, $\lambda^{-1}$ as well as its eigenvalues in any finite-dimensional block
with fixed $l$.

It is well known that for non-integer $I$ the eigenvalues of the second
$Q$-operator ${\bf A}^{(I)}_-(\lambda)$ are not polynomials. Having the explicit
form (\ref{qops16}) we can clarify this in details. The only non-analytic term in
$I$ in (\ref{qops16})
is the Pochhammer symbol in the numerator which can be transformed as follows
\beq
(\lambda^2q^{1-I+2(i'-n)};q^2)_{I-i-i'}=
(-\lambda^2 q^{i'-i-2n})^{I-i-i'}(\lambda^{-2} q^{1-I};q^2)_I
\frac{(\lambda^{-2}q^{1+I};q^2)_{n-i'}}
{(\lambda^{-2}q^{1-I};q^2)_{n+i}}.\label{qops23}
\eeq
So we see that for $I\in\mathbb{C}$ matrix elements of the operator
${\bf A}^{(I)}_-(\lambda)$ contain
a meromorphic function
\beq
(\lambda^{-2} q^{1-I};q^2)_I=\frac{(\lambda^{-2} q^{1-I};q^2)_\infty}
{(\lambda^{-2} q^{1+I};q^2)_\infty}\label{qops24}
\eeq
which doesn't depend on matrix indices. The rest of the formula (\ref{qops16}) is
the rational function of $q^I$ and can be analytically continued to  complex values
of $I\in\mathbb{C}$. As we have seen before in the case of the transfer-matrix,
both $Q$-operators for $I\in\mathbb{C}$ can be decomposed into an infinite
direct sum of finite-dimensional matrices with a fixed number of in- and out- spins.
Therefore, formulas (\ref{qops16}, \ref{qops22}) allow to construct
the $Q$-operators even in the case when the quantum space is the
tensor product of Verma modules $V_I^+$ with $I\in\mathbb{C}$.

Now let us consider some simple examples. First, take $I=1$.
Then we can write the operators ${\bf A}^{(I)}_\pm(\lambda)$ as 2-by-2
matrix operators acting in the auxiliary space of the $q$-oscillator
algebra (\ref{q-osc1}). In the representation of the $q$-oscillator algebra defined
by (\ref{conv6}) we obtain

\setlength{\extrarowheight}{5pt}

\beq
A^{(1)}_+(\lambda)=\phi^{-2\bosn}\begin{pmatrix} q^\bosn & -q^{-1}\bos^-\\
\lambda^{-1}\bos^+&
[\lambda^{-1}q^{\bosn}] \end{pmatrix},\label{qops25a}
\eeq
\beq
A^{(1)}_-(\lambda)=\phi^{2\bosn}\begin{pmatrix} [\lambda^{-1}q^{\bosn}]
 &\lambda^{-1}\bos^+ \\-q^{-1}\bos^- & q^\bosn \end{pmatrix},\label{qops25}
\eeq
\setlength{\extrarowheight}{6pt}
where $[x]$ is defined in $(\ref{conv2})$.

Similarly, one can calculate from (\ref{qops22}) the matrix elements
of $A^{(I)}_+(\lambda)$ at $I=2$ and obtain
\beq
A^{(2)}_+(\lambda)=\phi^{-2\bosn}
\begin{pmatrix}
q^{2\bosn}& -\bos^- q^{\bosn-1}_{\phantom{N}} & q^{-2}{(\bos^-)}^2 \\
 \lambda^{-1}\{q\}\bos^+ q^\bosn& \ds
 {\lambda^{-1}(\{q\}q^{2\bosn}-q^{-1})-\lambda}  &
q^{-\frac{5}{2}}\{q\}\bos^-\big[\lambda q^{-\bosn+\frac{1}{2}}\big]
  \\
\lambda^{-2}(\bos^+)^2& -q^{\frac{1}{2}}
\lambda^{-1}\bos^+[\lambda q^{-\bosn-\frac{1}{2}}] &
[\lambda q^{-\bosn-\frac{1}{2}}][\lambda q^{-\bosn+\frac{1}{2}}]
\end{pmatrix}.\label{qops26}
\eeq
The second L-operator $A^{(2)}_-(\lambda)$ is simply obtained from (\ref{qops26})
by reflection along rows and columns and changing $\phi\to\phi^{-1}$.

Now let us calculate the Wronskian $\text{\bf Wr}(\phi)$ in (\ref{qops12c}).
We do it by considering the limit $\lambda\to\infty$ and restricting (\ref{qops12c})
to the $l$-th sector where  the Wronskian is proportional to the identity matrix.

We start with calculating the limit of ${\bf A}^{(I)}_+(\lambda)$ at $\lambda\to\infty$.
It is easy to see from (\ref{qops22}) that  matrix elements of
${ A}^{(I)}_+(\lambda)$ at $\lambda\to\infty$ behave like

\setlength{\extrarowheight}{0pt}

\beq
[A^{(I)}_+(\lambda)]_{n,i}^{n',i'} \sim
\begin{cases}
\lambda^{2i'-i}(1+O(\lambda^{-2})), &\text{ for $i>i'$}\\
\phantom{aai}\lambda^i(1+O(\lambda^{-2})), &\text{ for $i\leq i'$ .}
\end{cases}\label{qops27}
\eeq
Therefore,  only the matrix
elements with the indices $i_k\leq i_k'$, $k=1,
\ldots,M$ will contribute to the leading order in $\lambda$.
It follows from (\ref{qops3}) that we need to consider
only diagonal matrix elements of ${\bf A}^{(I)}_+(\lambda)$.
Evaluating (\ref{qops22}) at $i'=i$ and taking the normalized trace
(\ref{conv8}) over
the auxiliary space
$\mathcal{F}_q$ in the $l$-th sector we get
\beq
{\bf A}_+^{(I)}(\lambda)|_{\lambda\to\infty} =
-(-\lambda)^l\phi^{2M}q^{2l-IM}({\boldsymbol I}+O(\lambda^{-2})),\label{qops28}
\eeq
where ${\boldsymbol I}$ is the identity matrix
and the trace in the denominator of the RHS of (\ref{conv8}) is equal to
\beq
{\underset{{\,\mathcal{F}_q}}{{\mbox{Tr}}}\,
({\phi^{2{\boldsymbol N}} q^{-{{\boldsymbol N}}\otimes H}})}=
\frac{1}{1-\phi^{2M}q^{2l-I M}}.\label{qops29}
\eeq
The asymptotics (\ref{qops28}) justifies our assumption (\ref{qops12})  made
to calculate the operator ${\bf A}_-^{(I)}(\lambda)$.

Taking the limit $\lambda\to \infty$ in (\ref{qops16}) we obtain by
the similar arguments
the asymptotics of the second $Q$-operator ${\bf A}_-^{(I)}(\lambda)$
\beq
{\bf A}_-^{(I)}(\lambda)|_{\lambda\to\infty} =
(-\lambda)^{I M-l}({\boldsymbol I}+O(\lambda^{-2})).\label{qops30}
\eeq

Substituting (\ref{qops4}, \ref{qops28}, \ref{qops30}) into (\ref{qops12c}) we obtain
the expression for the Wronskian in the $l$-th sector
\beq
\text{\bf Wr}^{(l)}(\phi)
=-(-1)^{I M}\phi^M q^{l-I M}(1-\phi^{2M} q^{2l-I M}) {\boldsymbol I}. \label{qops31}
\eeq
As expected the Wronskian does not depend on the spin $J$ in
the auxiliary space $V_J^+$.
We also notice that a normalization factor in (\ref{qops31}) depends on
the particular choice
of a $\lambda$-dependent normalization of the $L$-operators $A_\pm^{(I)}(\lambda)$.

\nsection{Functional relations}

In the previous section we used two fundamental
functional relations (\ref{qops6}) and (\ref{qops11}) which relate
transfer matrices ${\bf\widehat{T}}_{J,I}^{(l)}(\lambda;\phi)$,
${\bf {T}}_{J,I}^{(l)}(\lambda;\phi)$
and $Q$-operators ${\bf Q}_\pm^{(I)}(\lambda)$. Once they derived,
no further algebraic work is required.
All other functional relations are a consequence
of these two.
In this section we shall assume that $I\in\mathbb{Z}_+$ and
use operators ${\bf A}_\pm^{(I)}(\lambda)$ instead of ${\bf Q}_\pm^{(I)}(\lambda)$
since they are related by a simple transformation (\ref{qops12b}).

Since all the above operators commute, functional equations can be
rewritten in terms of its
eigenvalues. Let ${\widehat{\mathcal{T}}}_{J,I}^{(l)}(\lambda;\phi)$,
${\mathcal {T}}_{J,I}^{(l)}(\lambda;\phi)$ and ${\mathcal A}_\pm^{(I)}(\lambda)$
be the eigenvalues of the corresponding operators.

In the $l$-th sector of the quantum space we have
\beq
{\mathcal A}_+^{(I)}(\lambda)=\rho_+\prod_{k=1}^l [\lambda/\lambda_k^+],\quad
{\mathcal A}_-^{(I)}(\lambda)=
\rho_-\prod_{k=1}^{IM-l}[\lambda/\lambda_k^-].\label{func1}
\eeq

Let us start with the Wronskian relation between the eigenvalues of the
two $Q$-operators. Setting $J=0$ in
(\ref{qops6}) we obtain
\beq
\phi^{-M}{\mathcal A}_-^{(I)}(\lambda q^{1/2})
{\mathcal A}_+^{(I)}(\lambda q^{-1/2})-
\phi^M {\mathcal A}_+^{(I)}(\lambda q^{1/2})
{\mathcal A}_-^{(I)}(\lambda q^{-1/2})={\mathcal Wr}(\phi)
h_I(\lambda)^M,\label{func2}
\eeq
where ${\mathcal Wr}(\phi)$ denotes the eigenvalues of the Wronskian {\bf Wr}$(\phi)$
and $h_I(\lambda)$ is defined in (\ref{qops7}).

The functional equation (\ref{func2}) completely determines both polynomials
${\mathcal A}_\pm^{(I)}$
up to normalization factors $\rho_\pm$. Indeed, substituting into (\ref{func2})
$\lambda=\lambda_k^+q^{\pm1/2}$
and $\lambda=\lambda_k^-q^{\pm1/2}$ we obtain
\beq
\begin{split}
-\phi^M\mathcal{A}_+^{(I)}(q\lambda_k^+)\mathcal{A}_-^{(I)}
(\lambda_k^+)&={\mathcal Wr}(\phi)
h_I(\lambda_k^+ q^{1/2})^M,\\
\phi^{-M}\mathcal{A}_+^{(I)}(\lambda_k^+/q)\mathcal{A}_-^{(I)}
(\lambda_k^+)&={\mathcal Wr}(\phi)
h_I(\lambda_k^+ q^{-1/2})^M
\end{split}\label{func3}
\eeq
and
\beq
\begin{split}
-\phi^M\mathcal{A}_-^{(I)}(\lambda_k^-/q)\mathcal{A}_+^{(I)}
(\lambda_k^-)&={\mathcal Wr}(\phi)
h_I(\lambda_k^- q^{-1/2})^M,\\
\phi^{-M}\mathcal{A}_-^{(I)}(q\lambda_k^+)\mathcal{A}_+^{(I)}
(\lambda_k^-)&={\mathcal Wr}(\phi)
h_I(\lambda_k^- q^{+1/2})^M.
\end{split}\label{func4}
\eeq
From (\ref{func3}, \ref{func4}) we get two sets of Bethe ansatz equations
\beq
\phi^{\pm 2M}\frac{\mathcal{A}_\pm^{(I)}
(q\lambda_k^\pm)}{\mathcal{A}_\pm^{(I)}(q^{-1}\lambda_k^\pm)}=
-\left(\frac{h_I(\lambda_k^\pm q^{1/2})}
{h_I(\lambda_k^\pm q^{-1/2})}\right)^M.\label{func5}
\eeq
Of course, our derivation of the Bethe ansatz equations assumes that $\phi$ is
not equal to a special value when ${\mathcal Wr}(\phi)=0$.
We will not discuss here this and further subtleties like
the root of unity case $q^N=1$ (see \cite{BM07} for further discussions
and \cite{BLMS10} for the case $q=1$).

Combining (\ref{qops6}) and (\ref{qops11}) for an arbitrary $J$ we obtain
\beq
\begin{split}
h_{I-J}(\lambda)^M{\mathcal Wr}(\phi)&{\mathcal T}_{J,I}(\lambda;\phi)=\\
\phi^{-(J+1)M}&\mathcal{A}_+^{(I)}
(\lambda q^{-\frac{J+1}{2}})\mathcal{A}_-^{(I)}(\lambda q^{\frac{J+1}{2}})-
\phi^{(J+1)M}\mathcal{A}_+^{(I)}
(\lambda q^{\frac{J+1}{2}})\mathcal{A}_-^{(I)}(\lambda q^{-\frac{J+1}{2}})
\end{split}\label{func6}
\eeq
In particular, for $J=1$ we have
\beq
{\mathcal T}_{1,I}(\lambda;\phi)=
\frac{\phi^{-2M}\mathcal{A}_+^{(I)}(\lambda q^{-1})\mathcal{A}_-^{(I)}(\lambda q)-
\phi^{2M}\mathcal{A}_+^{(I)}(\lambda q)\mathcal{A}_-^{(I)}(\lambda q^{-1})}
{h_{I-1}(\lambda)^M{\mathcal Wr}(\phi)}
\label{func7}
\eeq
Multiplying (\ref{func7}) by $\mathcal{A}_\pm^{(I)}(\lambda $ and using
(\ref{func2}) we immediately arrive at the following equation
\beq
{\mathcal T}_{1,I}(\lambda;\phi){\mathcal A}_\pm^{(I)}(\lambda)=
\phi^{\pm M}\,[\lambda q^{\frac{1-I}{2}}]^M {\mathcal A}_\pm^{(I)}(q\lambda)+
\phi^{\mp M}\,[\lambda q^{\frac{1+I}{2}}]^M
{\mathcal A}_\pm^{(I)}(q^{-1}\lambda).\label{func8}
\eeq
We can rewrite (\ref{func8}) back in matrix form in terms of the original operators
${\bf Q}^{(I)}_\pm(\lambda)$. Using (\ref{qops12b})
we obtain the famous Baxter's $TQ$-relation
\beq
{\bf T}_{1,I}(\lambda;\phi){\bf Q}_\pm^{(I)}(\lambda)=
[\lambda q^{\frac{1-I}{2}}]^M {\bf Q}_\pm^{(I)}(q\lambda)+
[\lambda q^{\frac{1+I}{2}}]^M {\bf Q}_\pm^{(I)}(q^{-1}\lambda).\label{func9}
\eeq
We just proved that the operators ${\bf Q}^{(I)}_\pm(\lambda)$ are
two  solutions of (\ref{func9}).
Their linear independence has been proved earlier.

The $TQ$-relation (\ref{func8}) has a natural extension for any positive $J>1$. Combining
(\ref{func6}) with the Wronskian relation (\ref{func2}) one can show that

\beq
{\mathcal T}_{J,I}(\lambda;\phi)=\frac{{\mathcal A}_\pm^{(I)}(\lambda q^{-\frac{J+1}{2}})
{\mathcal A}_\pm^{(I)}(\lambda q^{\frac{J+1}{2}})}
{h_{I-J}(\lambda)^M}
\sum_{k=0}^J\frac{\phi^{\pm M(J-2k)}h_I(\lambda q^{k-\frac{J}{2}})^M}
{{\mathcal A}_\pm^{(I)}(\lambda q^{k-\frac{J+1}{2}})
{\mathcal A}_\pm^{(I)}(\lambda q^{k-\frac{J-1}{2}})}.\label{func10}
\eeq

Note that the formula (\ref{func10}) allows to express ${\mathcal T}_{J,I}(\lambda)$
in terms of the eigenvalues of only one $Q$-operator. So it can be more convenient
in cases when there is a problem to find the second linearly independent $Q$-operator.
In particular, (\ref{func9}) is useful in the limit $\phi\to1$ when two operators
${\bf A}_\pm^{(I)}(\lambda)$ can become linearly dependent and we can't apply
(\ref{func6}) due to the Wronskian being zero.

To conclude this section we shall give standard fusion relations satisfied by the eigenvalues
of the higher-spin transfer-matrices. For higher spin representations of the XXZ spin chain
they first appeared in
\cite{KR87a}. Since our normalization of the $R$-matrix is different, the scalar functions
in our formulas are modified comparing to \cite{KR87a}. We have
\beq
{\mathcal T}_{1,I}(\lambda;\phi){\mathcal T}_{J,I}(\lambda q^{\frac{J+1}{2}})=
f_{IJ}^{-}(\lambda q^{\frac{1-I}{2}})^M\,  {\mathcal T}_{J-1,I}(\lambda q^{\frac{J}{2}+1})+
f_{IJ}^{+}(\lambda q^{\frac{1+I}{2}})^M\, {\mathcal T}_{J+1,I}(\lambda q^{\frac{J}{2}}),
\label{func11}
\eeq
where
\setlength{\extrarowheight}{0pt}
\beq
f_{IJ}^+(\lambda)=
\begin{cases}
1,& \text{for $I>J$,}\\
[\lambda], &  \text{for $I\leq J$;}
\end{cases}\quad
f_{IJ}^-(\lambda)=
\begin{cases}
[\lambda][\lambda q^{I+1}],& \text{for $I\geq J$,}\\
[\lambda], &  \text{for $I<J$}
\end{cases}\label{func12}
\eeq
and
\beq
{\mathcal T}_{1,I}(\lambda){\mathcal T}_{J,I}(\lambda q^{-\frac{J+1}{2}})=
g_{IJ}^{-}(\lambda q^{\frac{I-1}{2}})^M\,
{\mathcal T}_{J-1,I}(\lambda q^{-\frac{J}{2}-1})+
g_{IJ}^{+}(\lambda q^{-\frac{1+I}{2}})^M\,
{\mathcal T}_{J+1,I}(\lambda q^{-\frac{J}{2}})\label{func13},
\eeq
\beq
g_{IJ}^+(\lambda)=
\begin{cases}
1,& \text{for $I>J$,}\\
[q\lambda], &  \text{for $I\leq J$;}
\end{cases}\quad
g_{IJ}^-(\lambda)=
\begin{cases}
[q\lambda][\lambda q^{-I}],& \text{for $I\geq J$,}\\
[q\lambda], &  \text{for $I<J$}.
\end{cases}\label{func14}
\eeq

We dropped a dependence on $\phi$ in the functional relations
(\ref{func12}, \ref{func14}) for brevity. The proof of these relations is similar.
We substitute the explicit expressions for the eigenvalues of the
transfer-matrices (\ref{func6}) and \eqref{func7} into
\eqref{func12}, \eqref{func14} and using the Wronskian relation
(\ref{func2}) reduce them to identity. All calculations are slightly tedious
but straightforward.

We also notice that all functional relations considered above can be
generalized to the case of complex $I\in\mathbb{C}$. The quantum space will be
infinite-dimensional $W=\stackrel[i=1]{M}{\otimes}V_I^+$ and decompose into
the direct sum of finite-dimensional blocks. All functional relations will
still be satisfied in any such block with properly
 modified scalar functions. We will leave this as the exercise
 for the reader.

\nsection{Conclusion}

In this paper we derived a new formula for the
$U_q(sl(2))$ $R$-matrix acting
in the tensor product of two highest weight modules with arbitrary weights
$I$ and $J$.
When $I=1$, this $R$-matrix reduces to the standard XXZ $L$-operator.
The formula for the matrix elements contains only one summation and
is expressed in terms of the basic hypergeometric series $\phifs$.
Taking the limit $I\to\infty$ we generalized the Bazhanov, Lukyanov and
Zamolodchikov
construction of  $Q$-operators to the XXZ spin chain with arbitrary spin.
This includes the infinite-dimensional case, when each $L$-operator acts in the
infinite-dimensional Verma module with a complex weight $J\in\mathbb{C}$.
These $Q$-operators are represented as special transfer-matrices
with an auxiliary space being the infinite-dimensional representation of
the $q$-oscillator algebra.
What is remarkable is that this construction is non-singular in the limit
$J\to \mathbb{Z}_+$.

However, as explained in the Introduction there is an alternative
construction of ``factorized'' $Q$-operators \cite{Der05,CDKK13} based on the
factorization property of the $U_q(sl(2))$ $L$-operator.
This approach works well for the infinite-dimensional representations
(or cyclic case $q^N=1$), but its restriction to a finite-dimensional case
requires a regularization.

The natural question now is how these two constructions of the
$Q$-operators are related to each other.
Since  both $Q$-operators commute with the transfer-matrix, there should be
a transformation between them which becomes singular in the limit of
integer weights.

Another interesting challenge is to construct the XXZ $Q$-operator
as the integral operator with a factorized kernel for the case of
infinite-dimensional
representations. This would allow to calculate the action of the $Q$-operator
on the proper (polynomial) basis directly and compare it with the results of
\cite{CDKK13}. We are going to address these problems in our next publication.

Let us remind
that a 3D approach of \cite{Bazhanov:2005as,Bazhanov:2008rd,MBS13} works
for the $U_q(\widehat{sl(n)})$ case with $n\geq2$. So it would be interesting to obtain
a generalization of the result (\ref{intro1}) for higher ranks.

Finally, we notice that the formula for the $R$-matrix  looks similar
to the expression of quantum $U_q(sl(2))$ 6j-symbols \cite{KR89}
in terms of the $q$-Racah polynomials \cite{AW79} (see also \cite{Rosen07}).
A better understanding of this connection and its possible generalization
to the elliptic case deserves a separate study.

\section*{Acknowledgments}
I would like to thank Sergei Derkachov, Jan De Gier, Gleb Kotousov  for
their interest to this work and useful discussions and Rinat Kashaev
for reading the manuscript and valuable comments.
I especially would like to thank Vladimir Bazhanov and Sergey Sergeev for
illuminating discussions, critical comments and
reading the manuscript.
This work is partially supported by the Australian Research Council.

\appendixtitleon
\begin{appendices}
\numberwithin{equation}{section}

\section{}
We start with the property
(\ref{zeroR}) which we reformulate as
\beq
R_{n_1,n_2,n_3}^{n'_1,n'_2,n'_3}=0 ,\quad \mbox{when}\quad
n_2>n_2',\quad -(n_2-n'_2)\leq n_3\leq -1. \label{B1}
\eeq
Using (\ref{rp1}) and (\ref{rp4}) we find that the matrix element in (\ref{B1}) is
proportional to
\beq
\phiss(q^{-2n_2},x;xq^{-2n'_2};q^2,z),\label{B2}
\eeq
where $x=q^{2+2n_1}$ and $z=q^{-2n_3}$. Using Heine's transformation of $\phis$
(see (III.3) in \cite{Gasper}) we obtain
\beq
\begin{split}
\phis(
q^{-2n_2},x;
xq^{-2n'_2}|q^2,z)=&\frac{
\left({q^2}/{z};q^2\right)_{n_2-n'_2}
\left(-{z}/{q^2}\right)^{n_2-n'_2}}{q^{(n_2-n'_2)(n_2-n'_2+1)}}\times\\
\times&\phis(q^{-2n'_2},xq^{2(n_2-n'_2)};xq^{-2n'_2};q^2,zq^{2(n'_2-n_2)}).
\label{B3}
\end{split}
\eeq
The factor $(q^2/z)_{n_2-n'_2}$ in the right hand side of (\ref{B3}) is equal
to zero for $z=q^{2k}$, $k=1,\ldots, n_2-n'_2$
which exactly corresponds to the range for the index $n_3$ from (\ref{B1}).

\section{}
Here we prove the identity between terminating balanced $\phifs$ series
which we used to prove (\ref{prop12}). Let us start with
the second Sears' transformation
\cite{Gasper}
\beq
\phifs\left(\left.
\begin{array}{l}q^{-m},a,b,c\\\phantom{aaaa}d,e,f\end{array}\right|q,q\right)=
\frac{(a,ef/ab,ef/ac;q)_m}{(e,f,ef/abc;q)_m}
\,\phifs\left(\left.\begin{array}
{l}q^{-m},q^{1-m}/d,e/a,f/a,\\q^{1-m}/a,ef/ab,ef/ac\end{array}\right|q,q\right)
\label{app2}
\eeq
where $def=abcq^{1-m}$, $m\in\mathbb{Z}_+$.

One can rewrite (\ref{app2}) in terms of regularized terminating series
(\ref{phi-reg}) as follows
\beq
\phif\left(\left.
\begin{array}{l}q^{-m};a,b,c\\\phantom{q^{-m},}d,e,f\end{array}\right|q,q\right)=
q^{m(m-1)}(ad)^m
\,\phif\left(\left.\begin{array}
{l}q^{-m};q^{1-m}/d,e/a,f/a,\\q^{1-m}/a,ef/ab,ef/ac\end{array}\right|q,q\right)\,.
\label{app3}
\eeq

Now let us choose $d=q^{1-m+n}$, where $m$ and $n$ are two positive integers and
apply the formula (\ref{app3}) first
with respect to the index $m$ and then with respect to the index $n$. After
simple transformations we get the following
result
\beq
\phif\left(\left.
\begin{array}{l}q^{-m};a,b,c\\q^{1-m+n},e,f\end{array}\right|q,q\right)=
\frac{(-1)^{m+n}(ab)^n (a,b,c;q)_{m-n}}
{q^{n+\frac{(m-n)(m-n-1)}{2}}}\ds\,\phif\left(\left.\begin{array}
{l}\ds q^{-n};\frac{q}{a},\frac{q}{b},cq^{m-n}\\ \ds q^{1-n+m},\frac{qe}{ab},\frac{qf}{ab}\end{array}\right|q,q\right),\label{app4}
\eeq
where
$abc=ef q^n $.

We also need another useful identity which allows to relate
matrix elements of two local $L$-operators
$A^{(J)}_+(\lambda)$ and $A^{(J)}_-(\lambda)$. It reads
\beq
\begin{split}
&(-1)^{j+j'}\frac{(q^2;q^2)_{J-j}(\lambda^2 q^{1-J-2(n-j')};q^2)_{J-j-j'}}
{(q^{2+2n};q^2)_{j-j'}(q^2;q^2)_j }\,\phit\left(\left.\begin{array}{l}
q^{-2j};q^{-2j'},\lambda^2q^{1-J}\\
q^{-2J},q^{2(1+n-j')}\end{array}\right|q^2,q^2\right)=\\
&=q^{j(j-1)-j'(j'-1)+2J(J-2j-n)+2n(j+j')}
\,\phit\left(\left.\begin{array}{l}
q^{-2(J-j)};q^{-2(J-j')},\lambda^2q^{1-J}\\
q^{-2J},q^{2(1+n-J+j)}\end{array}\right|q^2,q^2\right)\label{app5}
\end{split}
\eeq
provided that $0\leq j,j'\leq J$, $J\in \mathbb{Z}_+$ and $n, n+j-j'\geq0$.
This formula can be proved
by using transformations (III.9-III.13) in \cite{Gasper} for $\phit$ series.
We omit the details.

\section{}

In this appendix we derive the recurrence relations which completely determine
matrix elements $[R_{I,J}(\lam;1)]_{i,j}^{i'j'}$. Equation (\ref{prop15})
can be represented as a two-by-two matrix equation and leads to  four difference
equations for matrix elements. These four equations split into six equations
if we decouple them with respect to the spectral
parameter $\mu$. Further algebra shows that only three of them are linearly
independent provided that the indices satisfy the condition $i+j=i'+j'$.
Using the notation (\ref{prop16})
we obtain
\beq
\begin{split}
&(1-\lambda^2 q^{2(1+i+j)-I-J})(1-q^{2+2i'})S_{i,j}^{i',j'}-
\lambda q^{i'-j}(1-q^{2(1+i+i'-I)})(1-q^{2+2j})S_{i,j+1}^{i'+1,j'}\\
&-q^{3j-J-j'}
(1-q^{2+2i})(1-\lambda^2 q^{2(1+i'-j)+J-I})S_{i+1,j}^{i'+1,j'}=0,
\end{split}\label{Crecur1}
\eeq
\beq
\begin{split}
&\lambda q^{i-j-2i'+J-2}(1-q^{2(2+j+j'-J)})(1-q^{2+2i'})S_{i,j+1}^{i',j'+1}-
(1-\lambda^2 q^{2i-2j'+J-I})(1-q^{2+2j'})S_{i,j+1}^{i'+1,j'}\\
&+q^{3i-i'-I-1}(1-\lambda^2 q^{I+J-2(1+i+j)})(1-q^{4+2j})S_{i,j+2}^{i'+1,j'+1}=0,
\end{split}\label{Crecur2}
\eeq
\beq
\begin{split}
&\lam q^{3+3i-j-2I+J}(1-q^{2I-2i})(1-q^{2(J-j-j')})S_{i,j}^{i',j'}+
(1-q^{2(1+J-j)})(1-\lambda^2q^{2(1+i-j')+J-I})S_{i+1,j-1}^{i',j'}\\
&-q^{3+3i-i'-I}(1-q^{2J-2j'})(1-\lambda^2 q^{I+J-2i-2j})S_{i+1,j}^{i',j'+1}=0.
\end{split}\label{Crecur3}
\eeq
We can exclude shifts in two spins in the system (\ref{Crecur1}-\ref{Crecur3})
and derive a second order recurrence relation in one spin variable.
This recurrence relation is very similar to a recursion satisfied by q-Racah
polynomials. It has a unique solution which truncates for negative values of indices
and is given by (\ref{prop10}).

\end{appendices}

\bibliography{total32m}

\bibliographystyle{utphys}

\end{document}